\documentclass[letterpaper,aps,prd,reprint,flushbottom,preprintnumbers,nofootinbib,showkeys,superscriptaddress,longbibliography
]{revtex4-1}

\pdfoutput=1 % if your are submitting a pdflatex (i.e. if you have
\usepackage[T1]{fontenc}

\usepackage{amsfonts}
\usepackage{amsmath}
\allowdisplaybreaks
\usepackage{slashed}
\usepackage{amssymb}
\usepackage{latexsym}
\usepackage[dvipsnames]{xcolor}
\usepackage{float}
\usepackage{multirow}
\usepackage{graphicx}  % Add this in the preamble
\usepackage{lipsum} 
\usepackage[breaklinks]{hyperref}
\usepackage{enumitem}
\setlist[itemize]{topsep=4pt,itemsep=2pt,parsep=0pt,partopsep=0pt}
\hypersetup{colorlinks=true,citecolor=blue,linkcolor=blue,urlcolor=NavyBlue}
\usepackage[caption=false]{subfig}
\usepackage{natbib}
\usepackage{cancel}
\usepackage{relsize}
\usepackage{diagbox}
\usepackage{mathtools} % for definition :=
\usepackage{feynmp-auto} % for feynman diagram
\usepackage{bm}
\def\h1{\ensuremath{h_1}}
\def\h2{\ensuremath{h_2}}

\begin{document}
\title{
From CMB to LHC: A hybrid inflation model with gauged scale symmetry}
\author{Jinger Wu}
\email{wujinger22@mails.ucas.ac.cn}
\affiliation{School of Physics Sciences, University of Chinese Academy of Sciences, Beijing 100039, P.R. China.\\}

\author{Mei Huang}
\email{huangmei@ucas.ac.cn}
\affiliation{School of Nuclear Science and Technology, University of Chinese Academy of Sciences, Beijing 100039, P.R. China.\\}

\author{Qi-Shu Yan}
\email{yanqishu@ucas.ac.cn}
\affiliation{Center for Future High Energy Physics, Chinese Academy of Sciences, Beijing 100049, P.R. China.}
\affiliation{School of Physics Sciences, University of Chinese Academy of Sciences, Beijing 100039, P.R. China.\\}

\author{He-Xu Zhang}
\email{zhanghexu@ucas.ac.cn}
\affiliation{School of Nuclear Science and Technology, University of Chinese Academy of Sciences, Beijing 100039, P.R. China.\\}

\begin{abstract}
We propose a hybrid inflation model based on gauged scale symmetry, in which an axion-like field drives inflation while the Standard Model Higgs serves as the waterfall field.
During slow-roll inflation, the Higgs is trapped in the electroweak-symmetric vacuum. When the inflaton reaches a critical value, the Higgs acquires a nonzero vacuum expectation value and and triggers electroweak symmetry breaking (EWSB) through the waterfall dynamics.
We identify viable parameter regions that simultaneously accommodate current cosmic microwave background observations and Higgs data from the Large Hadron Collider. The relaxation mechanism connects the inflationary and electroweak scales by reducing the large vacuum-energy contribution required during inflation. Our framework thus provides a dynamical connection between inflation and EWSB.
\end{abstract}


\maketitle
\setlength{\parskip}{0pt}

\section{\label{sec:Intro}Introduction}
Inflation provides a compelling framework for addressing the horizon and flatness problems of the standard Big-Bang model~\cite{Guth:1980zm,Linde:1981mu,Albrecht:1982wi}. The primordial perturbations generated during inflation leave observable imprints on the cosmic microwave background (CMB)~\cite{Mukhanov:1981xt,Guth:1982ec,Planck:2018jri}. A wide variety of inflationary scenarios have been proposed~\cite{Starobinsky:1980te,Linde:1983gd,Freese:1990rb,Linde:1993cn,Bezrukov:2007ep,Kallosh:2013hoa}, among which models that connect inflationary dynamics to particle physics are of particular interest~\cite{Lyth:1998xn,Dvali:1994ms,Ballesteros:2016euj,
Ishida:2019wkd,Zhang:2023acu}.

A prominent example is Higgs inflation~\cite{Bezrukov:2007ep}, in which the Standard Model (SM) Higgs field itself plays the role of the inflaton. Its appeal lies in its minimal particle content, since no additional scalar inflaton beyond the SM is required. The nonminimal coupling of the Higgs field to gravity flattens the potential at large field values and yields robust predictions for the scalar spectral index $n_s$ and tensor-to-scalar ratio $r$~\cite{Bezrukov:2008ej}. Moreover, since the interactions of the Higgs with other SM particles are experimentally well constrained, the postinflationary reheating dynamics can be studied in considerable detail~\cite{Garcia-Bellido:2008ycs,Rubio:2018ogq}.

However, a nontrivial aspect of Higgs inflation is the connection between electroweak-scale Higgs physics and inflationary dynamics.
The couplings relevant for inflation are obtained by evolving those measured at the electroweak scale over many orders of magnitude in energy through renormalization-group (RG) equations~\cite{DeSimone:2008ei,Bezrukov:2009db,Allison:2013uaa,Burgess:2014lza}. Their values at the inflationary scale can therefore differ substantially from their low-energy values. Relating CMB observables to Large Hadron Collider (LHC) Higgs measurements consequently requires additional assumptions about the physics between the electroweak and inflationary scales~\cite{Bezrukov:2010jz,Bezrukov:2014bra,Bezrukov:2014ipa,He:2014ora, Xianyu:2014eba,Enckell:2016xse}.
Extensions based on gauged scale symmetry (Weyl symmetry)~\cite{Ghilencea:2018thl,Aoki:2022csb,Ghilencea:2021lpa,Wu:2025hfp} provide a broader inflationary framework, but simultaneously accommodating inflationary observables and collider constraints remains challenging.

In addition to this theoretical difficulty, recent CMB measurements have introduced a potentially interesting observational development. Some Atacama Cosmology Telescope (ACT)-inclusive data combinations, particularly those including baryon acoustic oscillation (BAO) measurements, favor a larger $n_s$ than Planck 2018 and can place conventional Higgs inflation under tension at the approximately $2\sigma$ level~\cite{AtacamaCosmologyTelescope:2025blo,AtacamaCosmologyTelescope:2025nti}.
Although this preference depends on the choice of data combination~\cite{Ferreira:2025lrd,McDonough:2025lzo,Balkenhol:2025wms}, it has stimulated renewed interest in inflationary realizations that yield a higher $n_s$ while maintaining a suppressed $r$~\cite{Yin:2025rrs,Yuennan:2025inm,Kallosh:2026qrc}.

One possible direction beyond conventional Higgs inflation is to consider a multiscalar framework in which the Higgs need not itself serve as the inflaton.
The additional degrees of freedom in multiscalar theories allow for richer inflationary dynamics and greater flexibility. A particularly simple realization is hybrid inflation~\cite{Linde:1993cn}, in which an inflaton is coupled to a waterfall field.
When the inflaton reaches a critical value, the waterfall field becomes unstable and develops a nonzero vacuum expectation value (VEV), thereby ending inflation.

This structure opens the possibility of identifying the waterfall field with the SM Higgs. Previous work~\cite{Garcia-Bellido:1999xos} explored such a realization in electroweak-scale hybrid inflation, with primary emphasis on postinflationary electroweak baryogenesis. However, radiative corrections strongly constrain hybrid inflation at such low scales and generally require the inflaton and/or the postinflationary trigger-field scale to lie well above the electroweak scale~\cite{Lyth:1999ty}. A straightforward realization of the SM Higgs as the waterfall field in low-scale hybrid inflation is therefore difficult.

Relaxation mechanisms have been proposed to dynamically adjust vacuum energies or generate hierarchically separated scales through cosmological evolution~\cite{Abbott:1984qf,Graham:2015cka,Graham:2019bfu}. We incorporate this idea into hybrid inflation by taking an axion-like field associated with a softly broken Peccei--Quinn (PQ) symmetry as the inflaton, while the SM Higgs serves as the waterfall field.
This allows the Higgs-waterfall mechanism to be realized in high-scale hybrid inflation while maintaining a connection to electroweak-scale physics.

In this work, we propose a hybrid inflation model with gauged scale symmetry, in which an axion-like field drives inflation while the SM Higgs serves as the waterfall field. When the inflaton reaches a critical value, the Higgs becomes unstable and acquires a nonzero VEV, triggering electroweak symmetry breaking (EWSB). The resulting Higgs dynamics provides the waterfall mechanism that terminates inflation, thereby establishing a direct connection between inflation and EWSB. We identify viable parameter regions that accommodate current CMB observations and Higgs measurements at the LHC.

The remainder of this paper is organized as follows. In Sec.~\ref{sec:Hybrid Model}, we introduce the guiding principles of the model construction and present the resulting framework. In Sec.~\ref{sec:Inflation}, we examine the viability of the inflationary scenario by determining the allowed parameter space and studying its theoretical consistency and phenomenological constraints. In Sec.~\ref{sec:EW}, we discuss the particle-physics implications of the model. We summarize our results and present the conclusions in Sec.~\ref{sec:summary}. Appendix~\ref{d.o.f.} is devoted to an analysis of the dynamical degrees of freedom in the gravity-scalar sector.

\section{\label{sec:Hybrid Model}A Hybrid Model with Weyl symmetry}
We extend the dilaton model proposed in Ref.~\cite{Wu:2025hfp} by introducing an additional real scalar field, denoted by $\Phi_2$. The original dilaton is denoted by $\Phi_1$, while $\phi$ denotes the SM Higgs doublet. It is natural to have such an extension in the context of Superstring~\cite{McAllister:2023vgy} and Supergravity~\cite{Kallosh:2002gf}, where such scalars could be moduli fields. When we fix $\Phi_1$ as a dilaton field, the new scalar field $\Phi_2$ can be either CP even or CP odd, which is subject to further theoretical assumptions.

Under the assumption with power terms and a gauged symmetry which is $ D(1) \otimes SU(2)_L \otimes U(1)_Y$, the Lagrangian can be put as 
\begin{widetext}
\begin{equation}
			\begin{aligned}
				\sqrt{-g} {\cal L} =& \sqrt{-g} \left( {\cal L}_k - {\cal L}_V \right )\,,\\
				{\cal L}_k =& \frac{\hat R^2}{4!\xi^2} -\frac{\hat C_{\mu\nu\rho\sigma}^2}{\eta^2}  - \frac{1}{4 g_s^2} \hat F_{\mu\nu}\hat F^{\mu\nu} - \frac{1}{4 g_w^2}  W_{\mu\nu}^a  W^{\mu\nu,a}+  \frac{1}{2} \sum_{i=1,2} \kappa_{ij} \hat \nabla_\mu \Phi_i \hat \nabla^\mu \Phi_j +  ( \hat \nabla_\mu \phi)^\dagger  \hat \nabla^\mu \phi   \,, \\
				 {\cal L}_V  = &
			\sum_{i,j,k,l=1,2} \frac{ \rho_{ijkl}}{4!} \Phi_i \Phi_j \Phi_k \Phi_l
			 + \sum_{i,j=1,2} \frac{\alpha_{ij}}{2}\Phi_i \Phi_j \phi^\dagger \phi + \frac{\lambda_0}{4} (\phi^\dagger \phi)^2 + \left [ \sum_{i=1,2} \frac{\beta_{ij}}{2}\Phi_i \Phi_j + \gamma \phi^\dagger \phi \right ] \hat R \,,
			\end{aligned}
			\label{LagarangianUV}
		\end{equation}
\end{widetext}
where we neglect the fermionic sector of the SM for the sake of simplicity. $\hat F_{\mu\nu}$ and $ W_{\mu\nu}^a$ are the field strength tensors for the Weyl vector and $SU(2)_L \otimes U(1)_Y$ gauge fields, respectively. The covariant derivative $\hat \nabla_\mu$ contains both the gauge fields and the Weyl vector. 

The couplings $g_s$ and $g_w$ represent the gauge couplings for the gauged scale symmetry and $SU(2)_L \otimes U(1)_Y$ symmetry, respectively. The parameters $\xi$, $\eta$, $\rho_{ijkl}$, $\alpha_{ij}$, $\beta_{ij}$, $\gamma$, $\lambda_0$ are all dimensionless couplings. 

The kinetic functions $\kappa_{ij}$ are metric of moduli space which typically depend upon moduli fields $\Phi_i$. Due to the existence of nonminimal couplings, the rank of the matrix $\kappa_{ij}$ should be consistent with the assumption of the number of dynamic degree of freedom, as in the BD theory $\omega \neq \frac{3}{2}$. We discuss this subtlety in detail in Appendix~\ref{d.o.f.}.

Here we would like to remark that except the gauge symmetry $SU(2)_L\times U(1)_Y$ of the SM, a gauged scale symmetry $D(1)$ is introduced, extending the Riemann geometry to the Weyl geometry. The gauged scale transformation of the fields and geometric quantities are given by,
\begin{equation}
\begin{aligned}
g_{\mu\nu} &\to \Omega^2 g_{\mu\nu}\,, \quad & g^{\mu\nu} &\to \Omega^{-2} g^{\mu\nu}\,, \\
\hat{R} &\to \Omega^{-2} \hat{R}\,, \quad & \hat{C}_{\mu\nu\rho\sigma} &\to \Omega^{-2} \hat{C}_{\mu\nu\rho\sigma}\,, \\
\Phi_i &\to \Omega^{-1} \Phi_i\,, \quad  & \phi  & \to \Omega^{-1} \phi\,,\\
\Psi &\to \Omega^{-3/2} \Psi\,, \quad &
\omega_\mu &\to \omega_\mu - \partial_\mu \ln\Omega\,,\\
\end{aligned}
\label{eq: ScaleTransformation}
\end{equation}
where $\hat R$ is the Weyl-invariant Ricci scalar, and $\hat C_{\mu\nu\rho\sigma}$ is the Weyl-invariant Weyl tensor. A real singlet scalar $\Phi_1$, the dilaton, is implemented in our model, as it is typically associated with scale symmetry. Both fermions $\Psi$ and the scalar doublet field $\phi$, which is the original Higgs doublet in SM, scale homogeneously, while the local gauge fields of the SM are invariant under the $D(1)$ symmetry. 

It should be noticed that the key element is the Weyl vector $\omega_\mu$, which is introduced by the gauged scale symmetry~\cite{hehl_metric_1995,Dirac:1973gk,trautman_geometry_1979} and thereby provides a bridge between Weyl geometry and Riemannian geometry,
\begin{equation}
\begin{aligned}
     \hat C_{\mu\nu\rho\sigma}^2 &= C_{\mu\nu\rho\sigma}^2 + 6 \,  F_{\mu\nu}F^{\mu \nu}\,,\\
 \hat R &= R - 6  (\omega_\mu \omega^\mu + \nabla_\mu \omega^\mu)\, ,\\
 \hat \nabla_\mu \Phi_i & =  \nabla_\mu \Phi_i - \omega_\mu \Phi_i\,, \\
  \hat \nabla_\mu \phi & = \nabla_\mu \phi - \omega_\mu \phi\,, \\
   \hat \nabla_\mu \Psi & = \nabla_\mu \phi - \frac{3}{2}\omega_\mu \Psi\,, 
  \label{W2R}
\end{aligned}
\end{equation}

Due to the constraint of $D(1)$ symmetry, in the potential part ${\cal L_V}$, there are 5 free parameters for the self-couplings $\rho_{ijkl}$, 3 free parameters for the singlet and doublet couplings $\alpha_{ij}$, and 3 free parameters for the nonminimal couplings $\beta_{ij}$. Thus in total, when the gauge couplings $\xi$, $\eta$, $g_s$, $g_w$, Higgs self-coupling $\lambda_0$ and Higgs nonminimal coupling $\gamma$ are also counted, the model has 17 free parameters. All these couplings are dimensionless, while those dimensional couplings are forbidden by the $D(1)$ symmetry.

Specifically, we assume that the dilaton $\Phi_1$ and new scalar $\Phi_2$ are singlets and are invariant under the EW symmetry group. Thus, the rest of gauge covariant differential operators are defined as 
\begin{equation}
	\begin{aligned}
		 \nabla_\mu \Phi_i &= \partial_\mu \Phi_i\,,\\
		 \nabla_\mu \phi &= (\partial_\mu - i T^a W_\mu^a) \phi\,,
	\end{aligned}
	\label{eq:the convariant derivative of scalars}
\end{equation}
where the $T^a$ and $W_\mu^a$ are the $SU(2)_L \otimes U(1)_Y$ generators and gauge fields, respectively.

In order to reduce the number of free parameters, we further impose a global symmetry to the Lagrangian given in Eq.~(\ref{LagarangianUV}), i.e. the PQ symmetry $U(1)_{PQ}$. The PQ symmetry is the essential ingredient of the relaxation mechanism, which provide a technical natural solution to the cosmological constant (C. C.) problem~\cite{Graham:2019bfu}. For the sake of convenience, we can combine $\Phi_1$ and $\Phi_2$ into a complex field $\Phi$. In order to realize the PQ symmetry, we can adopt either a linear form or a nonlinear form for the field $\Phi$,
\begin{equation}
\Phi =  \Phi_1+ i \Phi_2  = \frac{1}{\sqrt{2}} d e^{i \check{\theta}_a}\,,
\end{equation}
where $\check{\theta}_a=\frac{\theta_a}{f_A}$, and the $d$ denotes the dilaton field and $a$ denotes an axion-like field. The definition $\Phi=\Phi_1+ i \Phi_2$ is called  the Cartesian form, while $\Phi= \frac{1}{\sqrt{2}} d e^{i \check{\theta}_a}$ is called the polar form. Under the ansatz of PQ symmetry~\cite{peccei_cp_1977}, the fields in the model transform as given below
\begin{equation}
\begin{aligned}
g_{\mu\nu} &\to g_{\mu\nu}\,, \quad & g^{\mu\nu} &\to  g^{\mu\nu}\,, \\
\hat{R} &\to  \hat{R}\,, \quad & \hat{C}_{\mu\nu\rho\sigma} &\to \hat{C}_{\mu\nu\rho\sigma}\,, \\
\Phi &\to e^{i Q_{pq}} \Phi\,, \quad  & \phi  & \to e^{i Q_{pq}}  \phi\,,\\
\Psi &\to e^{- i \frac{\gamma_5}{2} Q_{pq}} \Psi\,, \quad &
\omega_\mu &\to \omega_\mu \,,\\
\end{aligned}
\label{eq: PQTransformation}
\end{equation}
where $Q_{pq}$ is the PQ charge. Thus the simplified Lagrangian can be organized as given below
\begin{widetext}
\begin{equation}
			\begin{aligned}
				\sqrt{-g} {\cal L} =& \sqrt{-g} \left( {\cal L}_k - {\cal L}_V \right )\,,\\
				{\cal L}_k =& \frac{\hat R^2}{4!\xi^2} -\frac{\hat C_{\mu\nu\rho\sigma}^2}{\eta^2}  - \frac{1}{4 g_s^2} \hat F_{\mu\nu}\hat F^{\mu\nu} - \frac{1}{4 g_w^2}  W_{\mu\nu}^a  W^{\mu\nu,a}+  (\hat \nabla_\mu \Phi)^* \hat \nabla^\mu \Phi+  ( \hat \nabla_\mu \phi)^\dagger  \hat \nabla^\mu \phi   \,, \\
				 {\cal L}_V  = &
			\frac{\rho_0}{4}(\Phi^* \Phi)^2
			 + \alpha \Phi^* \Phi \phi^\dagger \phi + \frac{\lambda_0}{4} (\phi^\dagger \phi)^2 + \left[ \beta \Phi^* \Phi + \gamma \phi^\dagger \phi \right ]  \hat R \,,
             \end{aligned}\label{LPQs}
			\end{equation} 
\end{widetext}
where we assume the metric of moduli space is flat and spacelike. Obviously, with the constraint of the PQ symmetry, the number of free parameters in the ${\cal L_V}$ is reduced to 5, and the total number of free parameters is only 9. 

This action is just has the same number of free parameters as the one studied in Ref.~\cite{Wu:2025hfp}, and it is difficult to cure the tension in order to accommodate CMB and Higgs data. In this work, we assume that the PQ symmetry is softly broken since the axion-like particle is assumed to be massive. For example, when only considering those renormalizable terms, we can introduce the following kinetic term
\begin{equation}
{\cal L}_k^b =  \frac{\delta_b}{2} ( \hat \nabla_\mu \Phi \hat \nabla^\mu \Phi+ \text{\text{h.c.}})    \,, \label{kpqb}
\end{equation}
which leads to a redefinition of the fields, $\Phi_1$ and $\Phi_2$, and consequently breaks the $U(1)$ symmetry. 
We can also introduce some terms in the potential, like the following terms 
\begin{equation}
\begin{aligned}
   {\cal L}_V^b  =   &\frac{\rho_1}{3!}(\Phi^* \Phi)(\Phi^2 + \text{h.c.})+  \frac{\rho_2}{4}(\Phi^2 + \text{h.c.})^2\\
   &+ \frac{ \alpha_1}{2}(\Phi^2 + \text{h.c.})\phi^\dagger \phi \,,\\
\end{aligned} \label{vpqb}
\end{equation}
Although these terms break the PQ symmetry, they respect a $Z_2 \times Z_2$ symmetry, i.e. they are invariant under the transformation $\Phi_1 \to \pm \Phi_1$ and $\Phi_2 \to \pm \Phi_2$ . It is found in our numerical analysis that these terms play the most important role in order to accommodate the CMB data.

While $\delta_b =0$ corresponds to the conventional PQ symmetric kinetic term commonly adopted in the literature, nonzero value of $\delta_b$, corresponding to PQ symmetry breaking in the kinetic sector, is also allowed by the symmetries of our model. 

We will consider the Lagrangian given in Eqs. (\ref{LPQs}-\ref{vpqb}) as the starting point of this study. Meanwhile, for the convenience of the following study, we set $\delta_b =1$ under the assumption that $\frac{\chi_H^2}{\chi_D^2} \neq \sqrt{1 + \frac{3 M_P^2}{v_{sm}^2}} - 1$, where $v_{\rm sm}$ denotes the electroweak vacuum expectation value in the SM. In the $\Phi_1$ and $\Phi_2$ form, by fixing $\delta_b=1$ the kinetic term of $(\partial \Phi_2)^2$ apparently vanishes. But due to the nonminimal couplings $\Phi_2^2 R$, $\Phi_2$ is still a dynamic field. The number of  degrees of freedom in such a setting are discussed in the Appendix~\ref{d.o.f.} in some details.

{The classic scaling symmetry $D(1)$ is broken when quantum corrections are taken into account. This breaking can be described in the context of strong dynamics using lattice simulations~\cite{Appelquist:2022mjb} or an effective field theory approach~\cite{Migdal:1982jp,Cata:2018wzl,Zwicky:2023krx}. In this study, we simply adopt the Stueckelberg mechanism to break the scaling symmetry.} 

Meanwhile, we adopt the unitary gauge for $\phi$ in the following analysis, which eliminates the Goldstone particles for the weak bosons,
\begin{equation}
	\phi = 
\begin{pmatrix}
	0 \\
	\frac{1}{\sqrt{2}}H \end{pmatrix}\,.
\end{equation}
Since both the dilaton field and the $H$ share the same quantum numbers $0^{++}$ ($J^{CP}$), it is expected that these two fields can mix with each other. Two physical fields can be produced after the mixing mechanism  determined by the nonminimal terms, such a mixing is very different from the diagonalization of mass matrix in the multiscalar models in particle physics. 

In order to guarantee that our Lagrangian can turn back to the general relativity (GR), it is observed that a Brans-Dicke (BD) field $\Theta$~\cite{Capozziello:2011et} can be defined, which is a combination of the scalar fields:
\begin{equation}
\Theta^2 \equiv \chi'^2_D \Phi^* \Phi  + \chi'^2_H H^2 \,,
\label{eq:dilaton_Higgs}
\end{equation}
where $\chi'_D$ and $\chi'_H$ are the mixing parameters for singlet field $\Phi$ and the doublet field $\phi$, respectively. After scale symmetry breaking, the BD field acquires a nonzero VEV, $\langle \Theta \rangle = f$, {thereby generating the Newton constant and the dimensional scales induced in the model.} It plays the role of a Goldstone field that is eaten by the Weyl vector and can be eliminated from the Lagrangian in the unitary gauge. 

{The remaining $0^{++}$ field is a Higgs scalar, which governs the EWSB, thereby providing a geometric origin for it~\cite{Flato:1987bb, deCesare:2016mml}.}

In the following study, we assume that $\chi_{D, H}' = \chi_{D,H} >0$, thus we can parameterize the scalar fields by the variables $\Theta$ and $\check \theta$ as:
\begin{equation}
    H^2 = \frac{\Theta^2}{\chi_H^2}   \sin^2\check{\theta}\,,\quad \Phi^* \Phi=\frac{\Theta^2}{\chi_D^2} \cos^2\check{\theta}\,. \label{case1}
\end{equation}
After using the unitarity gauge, i.e. $\omega_\mu \to \omega_\mu - \partial_\mu \frac{\Theta}{f}$, and substituting the parameterizations of the scalar fields given in Eq. (\ref{case1}), we obtain the relevant Lagrangian density for inflation. Since we focus on the inflationary epoch, the contributions from the Weyl tensor squared terms and the gauge fields terms can be neglected.

{ As pointed out in Ref.~\cite{Jarv:2016sow}, the inflationary observables depend solely on the terms of invariant potential, when kinetic terms of inflaton fields are defined in the canonical formalism in the effective field theory. It should be stressed that although in the whole framework the hybrid inflation intrinsically needs both inflaton and waterfall fields to work, during the slow-roll stage, the inflationary trajectory can be simply and effectively described by a single dynamical field since the waterfall field is trapped in the false vacuum and vanishes. 

So in order to find the inflationary observable and to cast the kinetic term into the canonical form, it is legitimate to replace the physical field $\check{\theta}_a$ by the auxiliary field $\xi$, which is identified as the inflaton in this work.}

Thus the inflationary Lagrangian density can have the following form 
\begin{equation}
	\begin{aligned}
		{\cal L} = &\frac{1}{2}\Big[(\nabla_\mu f_H \sin \frac{\theta}{f_H})^2+ (\nabla_\mu \xi)^2  \Big] - V(\theta,\xi),\\
		 V(\theta,\xi)  = & \begin{aligned}[t]
			&-\mu' (f_H \sin \frac{\theta}{f_H})^2 + \lambda' (f_H \sin \frac{\theta}{f_H})^4 \\
			&+ \rho' (f_H \sin \frac{\theta}{f_H})^2 \xi^2
			- \hat{\mu}'\xi^2 + \hat{\lambda}' \xi^4 + V_0,
		\end{aligned}
	\end{aligned}
	\label{L_inf}
\end{equation}
with the following short-hand definitions of fields 
\begin{equation}
	\begin{aligned}
		f_H &\equiv \frac{f}{\chi_H}\,, \quad f_D \equiv \sqrt{2} \frac{f}{\chi_D}\\
		\theta &\equiv  \check{\theta} f_H\,, \quad \xi \equiv f_D \cos \check{\theta} \cos \check{\theta}_a\,.
	\end{aligned}
	\label{eq:inflaton}
\end{equation}
And the parameters in Eq.~\eqref{L_inf} are determined from the original Lagrangian as:
\begin{equation}
	\begin{aligned}
		\mu' \equiv& f_D^2[\frac{\chi_H^2}{\chi_D^2}(  \frac{1}{4}\rho_0+ \rho_2-\frac{1}{3}  \rho_1)-\frac{1}{2}(\alpha -   \alpha_1)]\,,\\
		\lambda' \equiv& \frac{\lambda_0}{4} - ( \alpha -  \alpha_1)\frac{\chi_H^2}{\chi_D^2}+ ( \frac{1}{4}\rho_0+ \rho_2- \frac{1}{3}  \rho_1) \frac{\chi_H^4}{\chi_D^4}\,,\\
		\rho' \equiv & \alpha_1 +(2  \rho_2 -  \frac{1}{3}  \rho_1) \frac{\chi_H^2}{\chi_D^2}\,,\\
		 \hat{\mu}' \equiv& f_D^2 (  \rho_2 - \frac{ \rho_1}{3!}) \,,\\
		\hat{\lambda}'\equiv & \,\rho_2\,,\\
		V_0' \equiv &   f_D^4( \rho_0+4  \rho_2 - \frac{4}{3}  \rho_1)\,.
	\end{aligned}
	\label{eq:parameter relation}
\end{equation}
$V_0' = V_0 >0$ is assumed, which preserves the advantage of scale symmetry in generating a positive cosmological constant through its breaking. To arrive this potential, we have used the trigonometric identities $\sin^2(\check{\theta}_a) = 1 - \cos^2 (\check{\theta}_a)$.
About the Lagrangian given in Eq. (\ref{LagarangianUV}) and Eqs. (\ref{LPQs}-\ref{vpqb}), {it is worth emphasizing that, in quadratic gravity, renormalizability is ensured by the presence of higher-derivative curvature terms~\cite{Stelle:1976gc}, while gauged scale symmetry provides a natural organizing principle for such a structure.}
The introduction of the Weyl gauge field, and especially the replacement of the Einstein-Hilbert term by the nonminimal couplings, may cause a mixing among the Weyl vector, scalar fields, and the graviton, which leads to different propagator of graviton when compared with that in GR. {It is well-known that quadratic gravity is plagued with ghost issues. For this reason, we restrict to expose the graviton propagator in quadratic gravity and decompose the degrees of freedom of the theory.}
When necessary, it is possible to explicitly introduce a scalar field in the Lagrangian.
By doing so, it is noticed that the nonminimal couplings $\beta$ and $\gamma$ in Eq.~\eqref{LagarangianUV} can be modified into the couplings  $\chi_D'$ and $\chi_H'$ in Eq.~\eqref{eq:dilaton_Higgs}. 

\section{\label{sec:Inflation}Model Viability}
\subsection{The parameter regions for Inflation}
To realize the hybrid inflation scenario, we start from the effective potential in the Einstein frame, given by Eq.~\eqref{L_inf}. The first and second derivatives of the potential with respect to the inflaton $\xi$ and the waterfall field $\theta$, also identified as the Higgs field in our model, are given by:
\begin{align}
V_\theta &= \frac{\partial V}{\partial\theta} \notag \\*
 &= f_H\sin\frac{2\theta}{f_H}
 \Bigl[-\mu'+\rho'\xi^2+2\lambda'(f_H\sin\tfrac{\theta}{f_H})^2\Bigr]\,,\notag \\
V_{\theta\theta} &= \frac{\partial^2 V}{\partial\theta^2} \notag \\*
 &= 2\Bigl\{\cos\frac{2\theta}{f_H}
 \Bigl[-\mu'+\rho'\xi^2+2\lambda'(f_H\sin\tfrac{\theta}{f_H})^2\Bigr]\notag \\
 &\qquad +\lambda'(f_H\sin\tfrac{2\theta}{f_H})^2\Bigr\}\,,\notag \\
V_\xi &= \frac{\partial V}{\partial\xi} \notag \\*
 &= 2\xi\Bigl[\rho'(f_H\sin\tfrac{\theta}{f_H})^2-\hat\mu'
       +2\hat\lambda'\xi^2\Bigr]\,,\notag \\
V_{\xi\xi} &= \frac{\partial^2 V}{\partial\xi^2} \notag \\*
 &= 2\Bigl[\rho'(f_H\sin\tfrac{\theta}{f_H})^2-\hat\mu'
       +6\hat\lambda'\xi^2\Bigr]\,.
\label{deriv_V}
\end{align}

\begin{figure}[h]
    \centering
    \includegraphics[width=\linewidth]{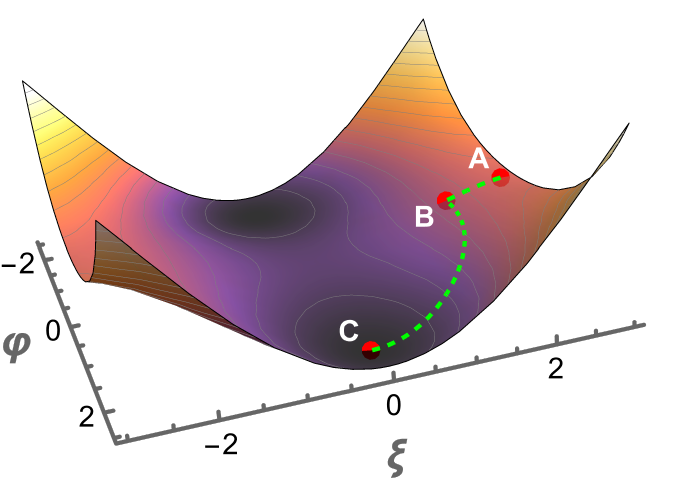}
    \caption{Schematic illustration of the potential in the hybrid inflation model. States $A$ and $B$ correspond to false vacua, while $C$ denotes the true vacuum of the Higgs field.}
    \label{fig:hybridInflation}
\end{figure}

It should be noticed that the signs of parameters of the effective potential are a priori undetermined, but they can be fixed by requiring a concave inflationary potential, a successful realization of hybrid inflation, and EWSB:
\begin{equation}
	\begin{aligned}
		\mu' \equiv \mu >0\,, \quad \lambda' \equiv \lambda >0\,, \quad \rho' \equiv \rho >0\,, \\
		\hat{\mu}' \equiv  - \hat{\mu} <0\,, \quad \hat{\lambda}' \equiv - \hat{\lambda} <0\,.
	\end{aligned}
\end{equation}
There are two vacua for the waterfall field in hybrid inflation. From Eq.~\eqref{deriv_V}, the false vacuum of Higgs field, the critical value of the inflaton $\xi_c$, together with the true  vacuum of the Higgs field and the inflaton field are given by:
\begin{equation}
	\begin{aligned}
		\text{False vacuum: }f_H \sin \frac{2 \theta}{f_H} =& 0\,,\\
		\xi_c^2 =& \frac{\mu}{\rho}\,, \\
		\text{True vacuum: } (f_H \sin \frac{ \theta}{f_H})^2 =& \frac{\mu}{2 \lambda} \equiv U^2\,,\\
		\xi^2 =& 0\,.
	\end{aligned}
	\label{critical}
\end{equation}
We define $\sqrt{\frac{\mu}{2 \lambda}}$ as $U$ for convenience in the following discussion. Since this work does not consider quantum corrections, $\mu \equiv \frac{(88.6424)^2}{2} {\rm GeV}^2$ and $\lambda \equiv0.129958$ are simply taken as determined from the LHC data fits~\cite{Wu:2025hfp}. The schematic diagram of hybrid inflation is shown in Fig.~\ref{fig:hybridInflation}, illustrating the potential in the $\xi$-$\varphi$ plane, where $\varphi$ is a redefinition of the Higgs field introduced later. Contour lines are shown in the figure in gray, where the purple region represents the low potential and the yellow region represents the high potential. Three points are labeled in the figure, representing three important states. $A$ is the state when the Higgs field is at the false vacuum; $B$ is the state when the inflaton field reaches to the critical value $\xi_c$ and triggers the Higgs field to roll down to the true vacuum; and $C$ is the state when both fields are at the true vacuum. 

When the Universe undergoes inflation in the slow-roll regime, the Higgs field is actually trapped in the false vacuum, which is shown by the curve connecting $A$ and $B$. At the end of inflation, the electroweak symmetry breaking is triggered and the Higgs field goes to its true vacuum, shown by the curve connecting $B$ and $C$.

The effective potential, when the waterfall field, i.e. Higgs field, is in its false vacuum, can be simply expressed as 
\begin{equation}
		V(\xi)  =  \hat{\mu}\xi^2 - \hat{\lambda} \xi^4 + V_0\,,
		\label{slow_roll}
\end{equation}
{which can be used to classify inflationary models~\cite{Jarv:2016sow}. With dynamical field $\xi$, this potential has a simple power-law form, while when expressed in the dynamical field $\check{\theta}_a$, the potential can have the following explicitly periodic form,
\begin{equation}
\begin{aligned}
		V(\check{\theta}_a) & =  C_0  +  C_2 \cos(2 \check{\theta}_a ) + C_4 \cos(4 \check{\theta}_a )  \,,\\ 
         C_0  & =  V_0 - \frac{2 \hat{\mu} f_D^2 - \hat{\lambda} f_D^4}{4} \,,\\
          C_2 &= \frac{  \hat{\mu} f_D^2 - \hat{\lambda} f_D^4 }{2}\,,\\
          C_4 &= \frac{ \hat{\lambda} f_D^4}{8} \,.
		\label{axionpotential}
\end{aligned}        
\end{equation}
Compared with the action of the natural inflation, there are two obvious differences which should be highlighted. 
\begin{itemize}
    \item In our model, when $\delta_b=0$, the kinematic term has the form $f_D^2 \partial_\mu \check{\theta}_a  \partial^\mu \check{\theta}_a $, which is the same as in the axion inflation model. When $\delta_b=1$, the kinematic term has a form $f_D^2 \cos^2(\check{\theta}_a)\partial_\mu \check{\theta}_a  \partial^\mu \check{\theta}_a $, which is different from that of axion inflation model.
    \item In the natural axion inflation model, the potential takes a simple form $V = \Lambda^4 \left[ 1 + \cos(\frac{\phi}{f}) \right]$, which includes two free parameters $\Lambda$ and $f$ and is different from the one given in Eq. (\ref{axionpotential}), where more free parameters can enter into the potential. This explains why our prediction is different with that of natural model.

\end{itemize}

However, the equivalence induced by canonicalization should be understood as an equivalence of the single-field slow-roll inflationary dynamics, not necessarily as a full equivalence of the underlying theories. After the slow-roll regime, the axion-like degree of freedom encoded in the inflaton can play a nontrivial role. This is central to the relaxation mechanism for addressing the C. C. problem, which will be discussed in detail later.}

Notice that the potential is unstable at large field values due to the negative quartic coupling of the inflaton. However, since the inflaton field value during inflation is sub-Planckian, we assume that the inflaton field at the pivot scale satisfies $\xi^2_* < \frac{\hat{\mu}}{2 \hat{\lambda}}< M_P^2$, which the unstability does not significantly affect the inflationary predictions. And $M_P$ is the reduced Planck mass. For the inflation can happen at the pivot scale, we require $V_{\xi \xi}<0$, which can be satisfied by assuming $\xi^2_* > \frac{1}{3} \frac{\hat{\mu}}{2 \hat{\lambda}}$.

The slow-roll parameters of our model are given by:
\begin{align}
				\epsilon|_{\sin\frac{2 \theta}{f_H} = 0} = &\frac{ M_P^2}{ 2}(\frac{V_\xi}{V})^2\notag \\*
				 = & 8 M_P^2\left[\frac{\xi (\frac{\hat{\mu}}{2 \hat{\lambda}} -\xi^2 )}{(\xi^2 - \frac{\hat{\mu}}{2 \hat{\lambda}})^2 - (\frac{V_0}{\hat{\lambda}} + \frac{\hat{\mu}^2}{ 4 \hat{\lambda}^2})}\right]^2 \, ,\notag \\
				 \eta|_{\sin\frac{2 \theta}{f_H} = 0} = & M_P^2\frac{V_{\xi \xi}}{V} \notag \\*
				 = & - 12  M_P^2 \frac{(\frac{\hat{\mu}}{2 \hat{\lambda}}- \xi^2) -  \frac{2}{3} \frac{\hat{\mu}}{2 \hat{\lambda}} }{  (\xi^2 - \frac{\hat{\mu}}{2 \hat{\lambda}})^2 - (\frac{V_0}{\hat{\lambda}} + \frac{\hat{\mu}^2}{ 4 \hat{\lambda}^2})} \,,\notag \\
				 \gamma|_{\sin\frac{2 \theta}{f_H} =0} = & M_P^4 \frac{V_{\xi \xi \xi} V_\xi}{ V^2} \notag \\*
				 = & - 96 M_P^4 \frac{\xi^2 ( \frac{\hat{\mu}}{2 \hat{\lambda}} - \xi^2)}{\left[(\xi^2 - \frac{\hat{\mu}}{2 \hat{\lambda}})^2 - (\frac{V_0}{\hat{\lambda}} + \frac{\hat{\mu}^2}{ 4 \hat{\lambda}^2})\right]^2}\,.
			\label{slow-roll}
		\end{align}
The $V_{\xi\,,\,\xi \xi}$ are the first and second derivative of the potential with respect to the inflaton $\xi$ as defined in Eq.~\eqref{deriv_V}, while $V_{\xi \xi \xi}$ is the third derivative of the potential. The slow-roll parameter associated with the fourth derivative should exist in theory, since $V_{\xi \xi \xi \xi} \neq 0$. However, we do not write this term explicitly, since it does not play a significant role in the subsequent discussion.

Prior to calculating the inflationary observables, the slow-roll conditions in our model should be examined. According to the slow-roll parameters defined in Eq.~\eqref{slow-roll} and the assumptions on $\xi_*^2$, we can get:
\begin{equation}
	\begin{aligned}
		\frac{\epsilon}{\eta^2} = & \frac{1}{18 M_P^2} \left[\frac{\xi (\frac{\hat{\mu}}{2 \hat{\lambda}} - \xi^2)}{  \xi^2- \frac{1}{3}\frac{\hat{\mu}}{2 \hat{\lambda}}}\right]^2 < 1\,,\\
		\left\lvert \frac{\gamma}{\eta^2}\right\rvert  =  & \frac{2}{3}\frac{\xi^2 (\frac{\hat{\mu}}{2 \hat{\lambda}} - \xi^2)}{  \left[\xi^2- \frac{1}{3}\frac{\hat{\mu}}{2 \hat{\lambda}}\right]^2} < 1\,.\\
	\end{aligned}
	\label{slow-roll approx}
\end{equation}
Thus the slow-roll conditions can be satisfied by assuming $\left\lvert \eta \right\rvert  \ll 1$.

Within the slow-roll approximation, the number of $e$-folds in our model is given by:
		\begin{align}
				N^{\rm inf} = & \ln \frac{a_\text{end}}{a} = \int_t^{t_\text{end}} H d t = \int_{\xi_e}^{\xi_*}\frac{1}{M_p} \frac{d \xi}{\sqrt{ 2 \epsilon(\xi)}} \notag \\
				= & \frac{1}{ 4 M_p^2}  \int_{\xi_e }^{\xi_* } d \xi \left\lvert \frac{(\xi^2 - \frac{\hat{\mu}}{2 \hat{\lambda}})^2 -(\frac{V_0}{\hat{\lambda}}  +\frac{\hat{\mu}^2}{ 4 \hat{\lambda}^2})}{\xi (\xi^2 - \frac{\hat{\mu}}{2 \hat{\lambda}})} \right\rvert \notag \\
				= & \begin{aligned}[t]
					\frac{1}{8 M_p^2} \Big[&(\xi_*^2 - \xi_e^2) - \frac{\hat{\mu}}{2 \hat{\lambda}} \ln \frac{\xi_*^2}{\xi_e^2} \\
					& - (\frac{V_0}{\hat{\lambda}} + \frac{\hat{\mu}^2}{4 \hat{\lambda}^2})\frac{1}{\frac{\hat{\mu}}{2 \hat{\lambda}}} \ln \left( \frac{  \frac{\hat{\mu}}{2 \hat{\lambda}}- \xi_e^2}{ \frac{\hat{\mu}}{2 \hat{\lambda}}- \xi_*^2}\frac{\xi_e^2}{\xi_*^2} \right)\Big]\,,
				\end{aligned} 
			\label{$e$-fold}
		\end{align}
where $\xi_*$ and $\xi_e$ are the values of the inflaton at the horizon exit and the end of inflation, respectively. The end of inflation is determined by the critical value of the inflaton, which triggers a real vacuum of the Higgs field, given by $\xi_e^2 = \xi_c^2 = \frac{\mu}{\rho}$ as shown in Eq.~\eqref{critical}. The third line of Eq.~\eqref{$e$-fold} is obtained by considering the assumption that $\xi^2 < \frac{\hat{\mu}}{2 \hat{\lambda}}$.

The inflationary observables of our model, including the scalar spectral index $n_s$, the spectrum of scalar and tensor perturbations $\mathcal{P}_{s,t}$, and the tensor-to-scalar ratio $r$, are evaluated up to the one-loop order~\cite{Stewart:1993bc, david_h_lyth_primordial_nodate} by
	\begin{align}
        n_s ={}& 1-6\epsilon+2\eta \notag \\*
        &+2\Bigl[-\frac{5+36C}{3}\epsilon^2+(8C-1)\epsilon\eta\notag \\*
        &\qquad +\frac{1}{3}\eta^2-\frac{3C-1}{3}\gamma^2\Bigr]\,,\notag \\
		\mathcal{P}_s = &   \left[1 - (2 C + 1) \epsilon + C \eta\right]^2 \frac{1}{24 \pi^2 M_P^4} \frac{V}{\epsilon}\,,\notag \\
		\mathcal{P}_t = &\left[1 + (C-1) \epsilon\right]^2 \frac{2}{3 \pi^2}\frac{V}{M_P^4}\,, \notag \\
		r = & \frac{\mathcal{P}_t}{\mathcal{P}_s} = \left[\frac{1+(C-1)\epsilon}{1-(2C + 1)\epsilon + C \eta}\right]^2 16 \epsilon \,,
	\end{align}
where $C \simeq - 0.73$ coming from the corrections. {Given that our parameter scan imposes only the condition $|\eta|<1$, rather than the stronger slow-roll requirement $|\eta|\ll 1$, loop corrections may be nonnegligible.}

\begin{table}[ht]
\centering
\begin{tabular}{|c|c|}
\hline
Parameter 
& ACT DR6 + Planck + BAO \\
\hline
$n_s$ 
& $0.9743 \pm 0.0034$ \\
\hline
$\ln(10^{10} \mathcal{P}_s)$ 
&  $3.060_{-0.012}^{+0.011}$ \\
\hline
$r_{0.05}$ 
& $r_{0.05} < 0.038$ (95\% CL) \\
\hline
\end{tabular}
\caption{Constraints on the inflationary observables from the latest combined experimental data~\cite{AtacamaCosmologyTelescope:2025blo, AtacamaCosmologyTelescope:2025nti}.}
\label{ACT}
\end{table}
\begin{table}[ht]
\centering
\setlength{\tabcolsep}{1pt}
\begin{tabular}{|c|c|c|}
\hline
Parameter & $N_{0.05}^{\rm inf}=56$ & $N_{re}=0$  \\
\hline
$\sqrt{\hat{\mu}}/M_p$ 
&$[5.8840,\, 13.3808] \times 10^{-6}$
&$[6.2023,\, 10.9441] \times 10^{-6}$\\
\hline
$\hat{\lambda}$ 
&$[6.2825,\, 7.0226] \times 10^{-10}$
&$[6.4970,\, 6.9330] \times 10^{-10}$\\
\hline
$\sqrt[4]{V_0}/M_p$ 
&$[3.5609,\, 5.3092] \times 10^{-3}$
&$[3.6456,\, 4.8046] \times 10^{-3}$\\
\hline
$\rho$ 
&$[1.5807,\, 3.2543] \times 10^{-11}$
&$[1.4758,\, 3.8946] \times 10^{-11}$\\
\hline
\end{tabular}
\caption{Parameter ranges that yield inflationary predictions consistent with the latest combined experimental data~\cite{AtacamaCosmologyTelescope:2025blo, AtacamaCosmologyTelescope:2025nti} at the 1$\sigma$ level.}
\label{paras}
\end{table}

From the CMB observations discussed above, we find
\begin{equation}
	\text{rank}(\frac{\partial (N, n_s, P_s, r)}{\partial (\hat{\mu}, \hat{\lambda}, V_0, \rho)}) = 4\,,
\end{equation}
implying that the parameters $\hat{\mu}$, $\hat{\lambda}$, $V_0$, and $\rho$ {may be} fixed by the experimental data. After scanning the parameter space, we find that the predictions of our model are consistent with the latest combined experimental data (Table~\ref{ACT})~\cite{AtacamaCosmologyTelescope:2025blo, AtacamaCosmologyTelescope:2025nti} at the 1 $\sigma$ level, with the parameters chosen as shown in Table~\ref{paras}. The first column corresponds to the case with $N_{0.05}^{\rm inf}=56$ $e$-folds, while the second column corresponds to the case with no reheating epoch ($N_{re}=0$) with the relation between $N_k^{\rm inf}$ and $N_{re}$ given by~\cite{Dai:2014jja,Cook:2015vqa}:
\begin{equation}
	\begin{aligned}
		N_{re} = -\frac{4}{1-3w_{re}}\Big[N_k^{\rm inf} &+ \log \frac{\rho_e^{\frac{1}{4}}}{H_k^{\rm inf}}  + \log \frac{k}{a_0 T_0}  \\
		&+ \frac{1}{4}\log \frac{30}{\pi^2 g_{re}}  + \frac{1}{3}\log \frac{g_{re}^s}{g_0^s} \Big]\,.
	\end{aligned}
\end{equation}
Here, $w_{re}$ is the equation of state parameter during reheating, $N_k^{\rm inf}$ is the number of $e$-folds during inflation, $\rho_e$ is the energy density at the end of inflation, $H_k^{\rm inf}$ is the Hubble parameter at the horizon exit, $k$ is the comoving wavenumber, $a_0$ and $T_0$ are the present scale factor and temperature of the Universe, respectively. $g_{re}$ and $g_{re}^s$ are the effective number of relativistic degrees of freedom for energy density and entropy density at the end of reheating, respectively, while $g_0^s$ is the effective number of relativistic degrees of freedom for entropy density at present. In our analysis, we set $w_{re} = 0$, $k=0.05 \text{ Mpc}^{-1}$, $g_{re} = g_{re}^s = 106.75$, and $g_0^s = 3.38$.

\begin{figure}[t]
    \centering
    \includegraphics[width=\linewidth]{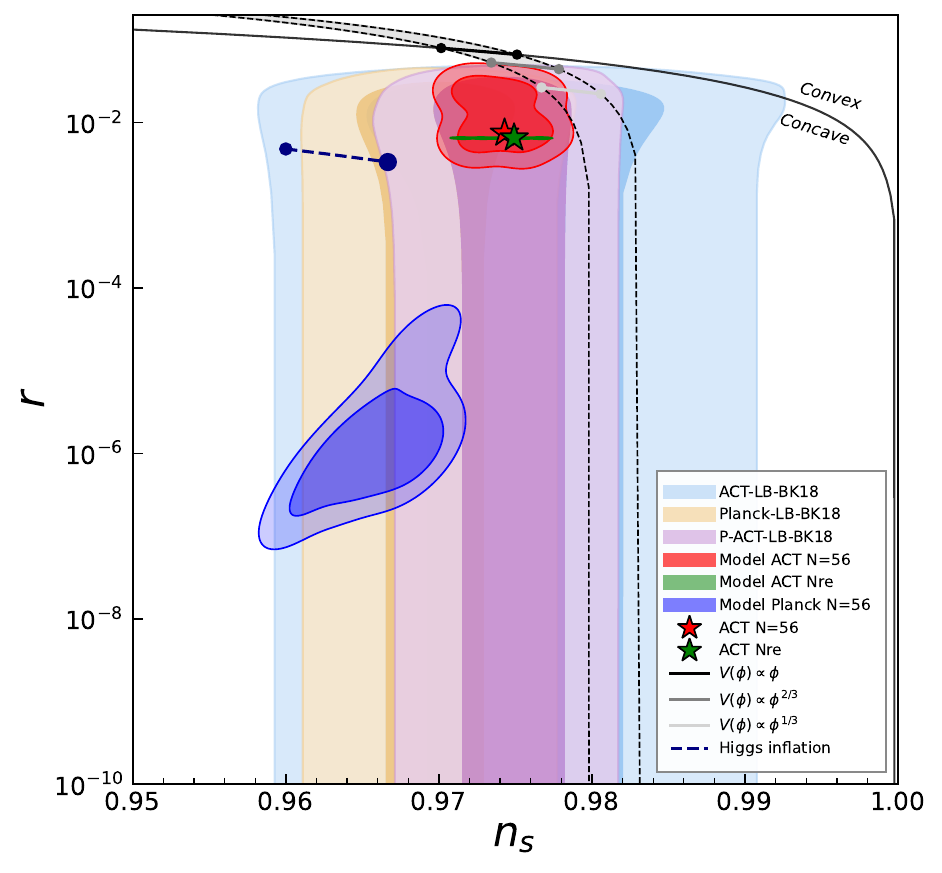}
    \caption{The $n_s-r$ plot on a logarithmic scale. We have used the MC data from~\cite{AtacamaCosmologyTelescope:2025blo, AtacamaCosmologyTelescope:2025nti} to plot the $1\sigma$ and $2\sigma$ contours for the datasets Planck2018 + CMB lensing + BK18 (orange), ACT DR6 + CMB lensing + BK18 (blue), and Planck2018 + ACT DR6 + CMB lensing + BK18 (purple). Inflationary potentials, such as power-law potentials and Higgs inflation, are also plotted with smaller points for $N=50$ and larger points for $N=60$. }
    \label{fig:ns-r}
\end{figure}

The values of $n_s$ and $r$ calculated using the parameters in Table~\ref{paras} are plotted in the $n_s$-$r$ plane in Fig.~\ref{fig:ns-r}, where the 1$\sigma$ and 2$\sigma$ contours are shown. The predictions for $N_{0.05}^{\rm inf}=56$ $e$-folds and $N_{re} = 0$ are shown in red and green, respectively. We also plot the region corresponding to $N_{0.05}^{\rm inf}=56$ $e$-folds for parameters constrained by Planck 2018 data (indigo). The star markers represent the predictions of $n_{s}$ and $r$ calculated using the representative benchmark points of the parameters in Table~\ref{paras}, with $N_{0.05}^{\rm inf}=56$ $e$-folds and $N_{re} = 0$ shown in red and green, respectively. The $1\sigma$ and $2\sigma$ contours of the experimental datasets~\cite{AtacamaCosmologyTelescope:2025blo, AtacamaCosmologyTelescope:2025nti} and the predictions of some inflationary potentials are also plotted for comparison. This shows that parameters constrained by the experimental data~\cite{AtacamaCosmologyTelescope:2025blo, AtacamaCosmologyTelescope:2025nti} indeed yield predictions of $n_s$ and $r$ consistent with the observations. Although the parameter space is not stable under experimental data, what we want to emphasize is the existence of such a parameter space, which is a nontrivial result given the tight constraints on $n_s$ and $r$.

\subsection{Additional Observational Constraints}
{Although the parameter scan identifies regions in which the required conditions are simultaneously satisfied, as shown in Table~\ref{paras}, further analyses of model consistency and phenomenological constraints are required.}

{A consistency check is required for the treatment of the $e$-folding number. In our parameter scan, the total $e$-folding number is assumed to be dominated by the slow-roll stage.
When $\xi^2$ reaches its critical value, the system subsequently undergoes a waterfall transition from the false vacuum to the true vacuum.For consistency, we assume that the waterfall transition contributes negligibly to the total number of $e$-folds. This assumption requires a self-consistency check within the adopted effective description.}

{To assess the plausibility of this assumption, we provide a rough estimate of the waterfall transition duration within the effective description. Although a full treatment requires solving the coupled evolution equations for $\theta_a$ and $\theta$, we use an effective description in terms of $\xi \equiv f_D\cos\check{\theta}\cos\check{\theta}_a$ and $\varphi \equiv f_H\sin\check{\theta}$ as a working approximation to obtain an order-of-magnitude estimate of the waterfall duration. We assume that this description remains approximately valid over the short interval of a rapid waterfall transition.}
Then, the evolution equations of the inflaton $\xi$ and the Higgs field $\theta$ can be derived from the effective Lagrangian in Eq.~\eqref{L_inf},
\begin{equation}
	\begin{aligned}
		\ddot{\xi} + 3 H \dot{\xi} =& -2 \xi \left[\rho \varphi^2 + 2 \hat{\lambda} (\frac{\hat{\mu}}{2 \hat{\lambda}} - \xi^2)\right]\,,\\
		\ddot{\varphi} + 3 H \dot{\varphi} =& -2 \varphi \left[\rho (\xi^2 - \frac{\mu}{\rho}) + 2 \lambda \varphi^2\right]\,.
	\end{aligned}
	\label{evolution}
\end{equation}
Under the assumptions $3 H \dot{\xi} \ll \ddot{\xi}$ and $3 H \dot{\varphi} \ll \ddot{\varphi}$, the evolution equations of $\xi$ and $\varphi$ admit approximate solutions given by
\begin{widetext}
	\begin{equation}
	\begin{aligned}
		\xi(t) &= \sqrt{(\rho \varphi^2 + \frac{\hat{\mu}}{2 \hat{\lambda}}) (1 - \sqrt{1 - 2 \hat{C}_1})} \text{sn}(\sqrt{\frac{4 \hat{\lambda} (\rho \varphi^2 + \frac{\hat{\mu}}{2 \hat{\lambda}})}{\sqrt{2}}(\sqrt{1 - 2 \hat{C}_1} + 1)}(t + \hat{C}_2), - \frac{1}{\hat{C}_1} (\hat{C}_1 + \sqrt{1 - 2 \hat{C}_1} - 1))\,,\\
		\varphi(t) &=  \sqrt{\frac{\rho}{2 \lambda}(\frac{\mu}{\rho} - \xi^2) (1 + \sqrt{1 + 2 C_1})} \text{sn}(\sqrt{\frac{4 \lambda \frac{\rho}{2 \lambda}(\frac{\mu}{\rho} - \xi^2)}{\sqrt{2}}(\sqrt{1 + 2 C_1} - 1)}(t + C_2), - \frac{1}{C_1} (C_1 + \sqrt{1 + 2 C_1} + 1))\,.
	\end{aligned}
	\label{approxsols}
\end{equation}
\end{widetext}
Here sn denotes the Jacobi elliptic function, while $\hat{C}_1$, $\hat{C}_2$ are integration constants
\begin{equation}
	\begin{aligned}
		& \hat{C}_1 = \begin{cases}
    \frac{\mu}{\rho} \frac{2 \hat{\lambda}}{\hat{\mu}}(\rho - \frac{\mu}{\hat{\mu}}\hat{\lambda}) \,,& \text{periodic type}\\
    \frac{1}{2} \,,  & \text{tanh type}
\end{cases}\\
		& \hat{C}_2 = -t_f\,,\\
	\end{aligned}
	\label{constants}
\end{equation}
fixed by the boundary conditions of $\xi$ at the critical point $t_c$ and at the end of the waterfall transition $t_e$. The first solution for $\hat{C}_1$ is obtained by assuming that $\xi^2$ corresponds to the maximum of the Jacobi elliptic function sn at the critical point, while the second solution corresponds to another branch of sn that gives the minimal duration of the waterfall transition.

The approximate solutions in Eq.~\eqref{approxsols} and also Eq.~\eqref{constants} are derived by neglecting the backreaction from the coupling between the inflaton and the Higgs field, which is justified for small $\rho$. Since $\varphi^2(t_c) = 0$ is certain from Eq.~\eqref{approxsols}, the evolution equation of $\varphi$ is difficult to determine. Thus, we reasonably assume that the waterfall duration for both $\varphi$ and $\xi$ is approximately equal. Then, the numerical results using the parameters of Table~\ref{paras} show that the waterfall transition duration $\Delta t = t_e - t_c$ can be estimated from Eq.~\eqref{constants} as $\Delta t \sim  t_H$ for the periodic type solution, while $\Delta t \ll t_H$ for the tanh type solution, where $t_H$ is the Hubble time at the critical point. The tanh type solution corresponds to the fast waterfall transition, which is the typical case in the hybrid inflation scenario. The periodic type solution corresponds to a slow waterfall transition, which can lead to a second phase of inflation during the waterfall process. However, with the slowest transition in our model lasting only $ t_H$, the process remains compatible with the CMB predictions presented above. Therefore, hybrid inflation in our model behaves as expected.

Compared with Higgs inflation, our model includes an additional waterfall field $\theta$, making it a multi-field inflation model, which can cause dynamics distinct from single-field models, such as the generation of isocurvature perturbations and non-Gaussianity. {These effects provide additional phenomenological constraints on the model.} We provide a rough approximation using the parameters listed in the first column of Table~\ref{paras} below.

According to the second line of Eq.~\eqref{deriv_V}, 
\begin{equation}
	V_{\theta \theta}(\frac{f_H}{2}\arcsin{0}, \xi)^{1/2} = \sqrt{2 (\rho \xi^2 - \mu)}\,,
\end{equation}
is lighter than the Hubble scale during inflation, indicating that entropy perturbations generated by the waterfall field exist, which may cause observable isocurvature perturbations. Although the exact solutions for $\xi$ and $\varphi$ are unknown, the approximate solutions in Eq.~\eqref{approxsols} can be used to estimate the order of magnitude of $\dot{\xi}$ and $\dot{\varphi}$,
\begin{equation}
	\frac{\dot{\varphi}}{\dot{\xi}} \sim \frac{\frac{\mu}{2 \lambda}(1 - \frac{\xi^2}{\xi_c^2})}{\frac{\hat{\mu}}{2 \hat{\lambda} }+ \rho \varphi^2} \sim 10^{-31}\,.
	\label{ev}
\end{equation} 
During the waterfall phase, the adiabatic and entropy fields can be approximately identified with the inflaton field and the Higgs field, respectively~\cite{Gordon:2000hv}. The change in the curvature perturbation $\mathcal{R}$ during the waterfall transition can be estimated as 
\begin{equation}
	\dot{\mathcal{R}} \propto  \frac{V_\theta}{\dot{\xi}} \sim \frac{\frac{\mu}{2 \lambda}}{\frac{\hat{\mu}}{2 \hat{\lambda} }} \sqrt{\frac{\mu}{2 \lambda}} \sim 10^{-29}\,.
\end{equation}
Combining with the approximate solutions for the duration of the waterfall transition $\Delta t$, the isocurvature perturbations generated by the entropy field during the waterfall transition are negligible, which is consistent with the latest CMB observations~\cite{AtacamaCosmologyTelescope:2025nti}.

The slow-roll process of hybrid inflation approximately follows the single-field slow-roll dynamics, we focus on the waterfall transition as the source of non-Gaussianity. Since we only consider the waterfall process here, we do not explicitly distinguish the $e$-folding number during waterfall process $N_{wf}$. For simplicity, we denote it by $N$. The $\delta N$ formalism~\cite{Sasaki:1995aw} can estimated the non-Gaussianity parameter $f_{\rm NL}$ as
\begin{equation}
		f_{\rm NL} = \frac{5}{6}\frac{N_{A} N_{B} N^{AB}}{(N_C N^C)^2} \,,
		\label{fNL}
\end{equation}
where $A$, $B$, and $C$ denote the fields $\xi$ and $\varphi$, {and the same indices means to sum over all}. Neglecting the cross-coupling effects since the isocurvature perturbations are negligible, we can obtain 
\begin{equation}
		\begin{aligned}
			N_\xi &\sim \frac{H}{\dot{\xi}}\,,\quad N_{\xi \xi} \sim -\frac{H \ddot{\xi}}{\dot{\xi}^3}\,,\\
			N_\varphi &\sim \frac{H}{\dot{\varphi}}\,,\quad N_{\varphi \varphi} \sim -\frac{H \ddot{\varphi}}{\dot{\varphi}^3}\,.
		\end{aligned}
\end{equation}
A similar estimation as in Eq.~\eqref{ev}, using Eq.~\eqref{approxsols}, gives 
\begin{equation}
	\ddot{\xi} \sim \left(\frac{\hat{\mu}}{2 \hat{\lambda}}\right)^{3/2}\,, \quad \ddot{\varphi} \sim \left(\frac{\mu}{2 \lambda}\right)^{3/2}\,.
\end{equation}
Therefore, the non-Gaussianity in our model is negligible since
\begin{equation}
	f_{\rm NL} \sim \frac{1}{H}\sqrt{\frac{\mu}{2 \lambda}} \sim 10^{-11}\,,
\end{equation}
which is consistent with the most stringent observational constraint $f_{\rm NL}^{\rm Planck} = -0.9 \pm 5.1$~\cite{Planck:2019kim}.

To briefly summarize the model presented above, we show that, using the CMB observational data in Table~\ref{ACT} and assuming that the waterfall field is identified with the SM Higgs field, the parameter space can be determined, which is also consistent with the constraints from isocurvature perturbations and non-Gaussianity.

\subsection{Model Viability Beyond Higgs Inflation}
{Before addressing the issues of the viability of our model, it is better to briefly review the viability of Higgs inflation, since our model includes the ingredient of it.}

The most salient feature of the Higgs inflation that makes it consistent with CMB~\cite{Planck:2018vyg} lies in the nonminimal coupling $\xi_h$ between the Higgs field and the Ricci scalar~\cite{Bezrukov:2007ep},
\begin{equation}
	\mathcal{L} \supset \xi_h h^2 R\,.
\end{equation}
After performing the transformation into the Einstein frame for inflationary analysis, the conformal factor $\Omega$, which is related to the nonminimal coupling constant through
\begin{equation}
	\Omega^2 \equiv 1 + \frac{h^2}{\Lambda^2}\,,
\end{equation}
where the cutoff scale is defined as $\Lambda \equiv \frac{M_P^2}{\xi_h}$, stretches the shape of the potential. In the regime where $\Lambda^2 > h^2$, {the Higgs potential is approximately the same as the original one}, while in the regime where $\Lambda^2 < h^2$, the potential becomes flat~\cite{Kallosh:2013hoa}, which is crucial for realizing inflation.
However, the conformal transformation not only stretches the potential but also modifies the kinetic term of the Higgs field, leading to a noncanonical kinetic structure,
\begin{equation}
	\begin{aligned}
		\mathcal{L}_{k} = \frac{1}{2}K(h)(\partial_\mu h)^2\,,\quad K(h) = \frac{\Omega^2 + 6 h^2/\Lambda^2}{\Omega^4}\,.
	\end{aligned}
\end{equation}
Because of the smallness of cutoff scale $\Lambda$, effective couplings such as $\frac{h^2}{\Lambda^2} (\partial_\mu h)^2$ can lead to unitarity violation at the scale $\Lambda$, known as the unitarity problem in Higgs inflation~\cite{Burgess:2009ea, Barbon:2009ya, Burgess:2010zq}.

Although similar nonminimal couplings also appear in our model, the key difference is that we work in the regime where $\Lambda^2 > h^2$. The introduction of scale symmetry permits the adoption of a large cosmological constant $V_0$ to obtain a flat potential, as shown in Table~\ref{paras}, which is the reason why this type of potential can successfully predict inflation. 
However, observations of the present Universe indicate that the cosmological constant is a tiny positive value~\cite{SupernovaSearchTeam:1998fmf,SupernovaCosmologyProject:1998vns} that conflicts with our model. This discrepancy is well-known as the C. C. problem, which is part of the price of circumventing the unitarity problem. This is also the reason why previous works cannot identify the waterfall field as the Higgs field. 

The model discussed above during inflation is based on the effective Lagrangian in Eq.~\eqref{L_inf}, which is derived from the Lagrangian in Eq.~\eqref{LagarangianUV} constrained by symmetries. 
{Although the relevant slow-roll trajectory is parametrized by the canonically normalized inflaton defined in Eq.~\eqref{eq:inflaton}, the underlying degree of freedom remains the axion-like field.}
A soft scale symmetry breaking term can be added in the potential of Eq.~\eqref{L_inf}, as in relaxation models~\cite{Abbott:1984qf},
\begin{equation}
	 \begin{aligned}
		{\cal L}^T_V  &= {\cal L}_V - k \theta_a \,.
	 \end{aligned}
	 \label{breaking}
\end{equation}
$k$ is a parameter with mass dimension 3. {Such a potential can be obtained from wrapped branes~\cite{McAllister:2008hb} straightforwardly.} Then, the derivatives of the potential with respect to the inflaton field $\xi$ in Eq.~\eqref{deriv_V} are modified by additional terms arising from the symmetry-breaking term:
\begin{equation}
	\begin{aligned}
		\frac{\partial \mathcal{L}_V^T}{\partial \xi} = & \frac{\partial \mathcal{L}_V}{\partial \xi}+ \frac{\partial (-k \theta_a)}{\partial \xi}  \sim \frac{\partial \mathcal{L}_V}{\partial \xi}\, ,\\
		\frac{\partial^2 \mathcal{L}^T_V}{\partial \xi^2} =& \frac{\partial^2 \mathcal{L}_V}{\partial \xi^2} +  \frac{\partial^2 (-k \theta_a)}{\partial \xi^2}\sim \frac{\partial^2 \mathcal{L}_V}{\partial \xi^2}  \, .\\
	\end{aligned}
	\label{mod_deriv}
\end{equation}

The last relation in above equations is obtained by considering the smallness of the parameter $k$, 
\begin{equation}
	\begin{aligned}
		\frac{\partial (-k \theta_a)}{\partial \xi}  = & \pm k \frac{f_A}{\sqrt{f_D^2 (1 - \frac{\varphi^2}{f_H^2}) - \xi^2}} \to 0 \,\\
	\end{aligned}
	\label{condition}
\end{equation}
which is crucial for preserving the inflationary predictions of our model. And this makes sense because the smallness of $k$ is associated with soft breaking.
After inflation, the system settles into the true vacuum, around which we expand the field as $\theta_a = v_a + a$, where $v_a$ is {the classical solution corresponding to this vacuum.} Then, Eq.~\eqref{critical} gives 
 \begin{equation}
		\frac{v_a}{f_A} = \arccos 0 = {\frac{\pi}{2} + N \pi\,.}
	\label{truevac}
 \end{equation}
The first equality of Eq.~\eqref{truevac} follows from $\xi^2 = 0$, while the second equality follows from the periodicity of trigonometric functions with $N \in \mathbb{Z}$, which is crucial for resolving the C. C. problem.

Before addressing the C. C. problem, it is useful to analyze the effective interactions in the low-energy regime,
\begin{equation}
	\begin{aligned}
		\mathcal{L}_{V}^T =& \begin{aligned}[t]
			&-\mu (f_H \sin \frac{\theta}{f_H})^2 + \lambda (f_H \sin \frac{\theta}{f_H})^4 \\
			&+ \rho (f_H \sin \frac{\theta}{f_H})^2 (f_D \cos \frac{\theta}{f_H} \cos \frac{\theta_a}{f_A})^2
			\\
			&+ \hat{\mu}(f_D \cos \frac{\theta}{f_H} \cos \frac{\theta_a}{f_A})^2 - \hat{\lambda} (f_D \cos \frac{\theta}{f_H} \cos \frac{\theta_a}{f_A})^4 \\
			&+ V_0 - k A \,,
		\end{aligned}\\
		 \Rightarrow &\begin{aligned}[t]
			&\frac{m_h^2}{2} h^2 + \kappa_{3h} v_{sm} \lambda_{sm} hhh + \kappa_{4h} \frac{\lambda_{sm}}{4}hhhh \\
			+&  \frac{m_a^2}{2} a^2+ \lambda_{1h2a} haa + \lambda_{2h2a} hhaa -k a+ \dots\\
			+& V_0 - \frac{\mu^2}{4 \lambda} - k v_a \,.
		\end{aligned}
	\end{aligned}
	\label{effective}
\end{equation}
The arrow denotes the effective potential after electroweak symmetry breaking, where the field $\theta$ is expanded as $\theta = v + h$, with the vacuum expectation value $v$ of $\theta$, and the field $A$ introduced earlier.

\begin{table}[ht]
\centering
\begin{tabular}{|l|c|}
\hline
$m_h^2$ 
& $4 \mu $ \\
\hline
$\kappa_{3h}$ 
&  $\frac{\left(1-2s^2\right)}{\left(1-s^2\right)}\frac{4 \lambda}{\lambda_{sm} }\frac{U }{v_{sm}}$ \\
\hline
$\kappa_{4h}$ 
& $\frac{\left(3 - 28s^2 + 28s^4\right)}{3  \left(1 - s^2\right)^2 \,}\frac{4 \lambda}{\lambda_{sm} }$ \\
\hline\hline
$m_a^2$ 
& $2 \left( \hat{\mu} +\rho U^2 \right)$ \\
\hline
$\lambda_{1h2a}$ & $2 \rho U - \frac{s^2}{1-s^2} \frac{m_a^2}{ U}$\\\hline
$\lambda_{2h2a}$ & $\frac{1-6s^2}{1-s^2}\rho-\frac{1}{2}\frac{s^2(1 - 2s^2)}{(1-s^2)^2}\frac{m_a^2}{U^2}$\\\hline
\end{tabular}
\caption{The explicit form of the effective potential corresponding to the experimental value is shown in Eq.~\eqref{effective}.}
\label{couplings}
\end{table}
The first line of the effective potential gives the SM contribution, including the mass term of $H_{125}$, the trilinear and quartic self-interactions of Higgs parametrized within the $\kappa$ framework. $\lambda_{sm}$ and $v_{sm}$ denote the Higgs self-coupling and the VEV of electroweak symmetry in the SM, respectively. The second line gives the contributions beyond the SM, including the mass term of the new scalar $a$, the trilinear interactions between $H_{125}$ and $a$, in particular the linear term in $a$ induced by the breaking term, as well as additional couplings denoted by the ellipsis. The cosmological constant is contained in the last line, which includes the contributions from scale symmetry, electroweak symmetry, and the explicit symmetry-breaking term introduced for relaxation. A parameter is introduced for simplicity,
\begin{equation}
	\begin{aligned}
		s &\equiv \sin \frac{v}{f_H}\,.
	\end{aligned}
\end{equation}
Then, the scalar mass terms $m_{h,\,a}$, kappa parameters $\kappa_{3h,,4h}$, and the effective couplings $\lambda_{1h2a}$ introduced above are summarized in Table~\ref{couplings}. Details of $m_h^2$ and $\kappa_{3h,\,4h}$ are discussed in Ref.~\cite{Wu:2025hfp}, with those of $m_a^2$ and $\lambda_{1h2a,\,2h2a}$ deferred to the next section.

From the last line of Eq.~\eqref{effective}, one can always obtain a solution of $v_a$ with
\begin{equation}
	 \begin{aligned}
		V_0 - \frac{\mu^2}{4 \lambda} - k v_a&=0\,,\\
		v_a = \frac{1}{k} (V_0 - \frac{\mu^2}{4 \lambda})\,.
	 \end{aligned}
	 \label{cc}
\end{equation}
Note that $v_a \gg f_A$ is an important property in the relaxation mechanism for solving the C. C. problem, which also ensures that the condition in Eq.~\eqref{condition} is not spoiled by $V_0$.
However, the explicit scale breaking term is unbounded, which is crucial for solving the C. C. problem, while it also causes vacuum instability of $a$.

If our Universe is currently in a metastable vacuum, the lifetime of the vacuum should be longer than the age of the Universe. Trivially, we obtain 
\begin{equation}
    \begin{alignedat}{4}
        \mathcal{L}_{V}(v,\, v_a) &\equiv \mathcal{L}_V^{\rm meta}\,, \\
        \mathcal{L}_{V}(v,\, v_a + \pi f_A) &\equiv \mathcal{L}_V^{\rm true}\,, \qquad
		&\Delta V &\equiv \mathcal{L}_V^{\rm meta} - \mathcal{L}_V^{\rm true}\,, \\
        \mathcal{L}_{V}(v,\, v_a + \tfrac{\pi}{2} f_A) &\equiv \mathcal{L}_V^{\rm max}\,, \qquad
        &\mathcal{E} &\equiv \mathcal{L}_V^{\rm max} - \mathcal{L}_V^{\rm meta}\,,
    \end{alignedat}
\end{equation}
from Eq.~\eqref{effective}, which satisfies the thin wall approximation $ \Delta V \ll \mathcal{E}$ when $k f_A/\text{GeV}^4 \ll 10^{31}$. Thus, the decay width of the metastable vacuum can be estimated as~\cite{Coleman:1977py}
\begin{equation}
	\begin{aligned}
		\frac{\Gamma}{V} &= A e^{-B/\hbar} (1 + \mathcal{O}(\hbar))\,,\\
		 B &= \frac{27 \pi^2 \sigma^4}{2 (\Delta V)^3}\,, \qquad A \sim \frac{1}{R_{c}^4}\,,
	\end{aligned}
\end{equation}
where $\sigma$ is the tension of the bubble wall and $R_c$ is the critical radius of the bubble, given by 
\begin{equation}
	\begin{aligned}
		\sigma &= \int_{v_a}^{v_a+ \pi f_A} d A \sqrt{2(\mathcal{L}_V - \mathcal{L}_{V}^{\rm true})}\,,\\
		R_c &= \frac{3 \sigma}{\Delta V}\,.
	\end{aligned}
\end{equation}
Using the parameters in the first column of Table~\ref{paras}, we numerically obtain $B \sim (\frac{10^{22}}{k f_A/\text{GeV}^4})^3$ and $R_c \sim \frac{10^{16}}{k f_A/\text{GeV}^4} \text{ GeV}^{-1}$, yielding the lifetime of the metastable vacuum 
\begin{equation}
	\tau = \frac{1}{\Gamma} \sim R_c\, e^{B}\sim \frac{10^{- 16}}{k f_A/\text{GeV}^4} e^{(\frac{10^{22}}{k f_A/\text{GeV}^4})^3} \mathrm{yr}\gg t_U \,,
	\label{lifetime}
\end{equation}
for $k f_A/\text{GeV}^4 \leq 10^{21} $, where $t_U$ is the age of our Universe. Thus, Eq.~\eqref{lifetime} implies that the metastable vacuum is effectively stable for the present Universe.

{Thus, the viability of our framework relies on two key ingredients. The gauged scale symmetry constrains the scalar potential and permits a large effective vacuum-energy contribution $V_0$, which is crucial for obtaining a sufficiently flat inflationary direction. At the same time, the relaxation dynamics dynamically reduces this additional vacuum-energy contribution after inflation, allowing the system to evolve toward the desired low-energy electroweak vacuum. The model retains the successful inflationary behavior of scale-symmetric constructions while avoiding a large residual cosmological constant that would otherwise obstruct the identification of the waterfall field with the Higgs sector.}

\section{\label{sec:EW}The phenomenology of particle physics}
Our model is motivated not only by the latest CMB observations but also by the desire to connect the physics of inflation to the electroweak scale. The coupling between the inflaton and the Higgs field allows the waterfall field to be identified as the Higgs field, which is a key feature of our model. Based on the dynamic connection, this section will briefly discuss the phenomenological implications of our model at the electroweak scale, which are also crucial for testing the model.

The coupling between the inflaton $\xi$ and the Higgs field $\theta$ allows different fields to coherently take over the dynamics at different stages. During inflation, the inflaton field $\xi$ dominates the dynamics, while the Higgs field $\theta$ is stabilized at the origin. The Higgs field $\theta$ is triggered to acquire a vacuum expectation value when the inflaton $\xi$ reaches its critical value, thereby spontaneously breaking the electroweak symmetry and generating the Higgs boson mass by
\begin{equation}
	m_h^2 = 4 \mu (1- \frac{ \xi^2}{ \xi_c^2})\left[ 1 - \frac{\frac{ \xi^2}{ \xi_c^2}}{2}\frac{3 - 4 s^2}{1 - s^2} \right]\,.
\end{equation}
The backreaction from the Higgs field $\theta$ also triggers the end of inflation, which is the typical feature of hybrid inflation. After the waterfall transition, the Higgs field governs the electroweak symmetry breaking dynamics while the inflaton field $\xi$ is stabilized at the origin.

An important issue concerns the robustness of the above results, as the Higgs boson mass is not protected by a symmetry in the same way as many other particle masses in the SM, leading to 
\begin{equation}
	m_h^2 = m_{h,\,0}^2 + \delta m_h^2\,,\quad \delta m_h^2 \sim \Lambda^2\,,
\end{equation}
where $m_{h,\,0}$ is the bare mass of Higgs boson, $\delta m_h$ is the radiative correction to the mass, $\Lambda$ is the cutoff scale of the theory. 

The model constructed in this work can alleviate the hierarchy problem. The dominant radiative corrections to the Higgs mass parameter arise from the couplings $\lambda_{1h2a}$ and $\lambda_{2h2a}$, 
\begin{equation}
		\delta m_h^2 \sim \frac{1}{16 \pi^2}(\lambda_{1h2a}^2 + \lambda_{2h2a} m_a^2)\,.
\end{equation}
Both $\lambda_{1h2a}$ and $\lambda_{2h2a}$ arise from the periodic structure shown in Eq.~\eqref{effective}, which follows from the underlying symmetries of the model. Thus, the condition
\begin{equation}
	\left\lvert (\frac{\lambda_{1h2a}}{m_a})^2 + \lambda_{2h2a}\right\rvert  \sim \frac{\delta m_h^2}{m_a^2} \lesssim \frac{m_h^2}{m_a^2}\,,
\end{equation}
can be realized when $\lambda_{2h2a}$ is negative and comparable in magnitude to $\lambda_{1h2a}^2/m_a^2$. 

Apart from the hierarchy problem, the model provides a dynamical way to address the vacuum stability problem. 
The conventional connection between the inflationary and electroweak scales relies on renormalisation-group running, which may lead to vacuum instability at high energy scales~\cite{Djouadi:2005gi,Degrassi:2012ry, Buttazzo:2013uya},
\begin{equation}
	\lambda(\mu_c) <0\,,\quad \mu_c \sim 10^{10} \text{ GeV}\,.
	\label{instability}
\end{equation}
In our numerical analysis, we directly use the electroweak scale parameters $\mu$ and $\lambda$ as inputs to explore the possibility of connecting the inflationary and electroweak scales without relying on renormalisation group running. In this sense, the model provides a useful handle on the viable parameter space, which dynamically preserves a positive effective quartic coupling in the EWSB region, thus evading the instability condition represented by Eq.~\eqref{instability}.

\section{\label{sec:summary}Summary and Discussion}
In this work, we have proposed a hybrid inflation scenario with gauged scale symmetry in which an axion-like field drives inflation while the SM Higgs field plays the role of the waterfall field. During the slow-roll stage, the Higgs is stabilized at the electroweak-symmetric vacuum. Once the inflaton reaches the critical value $\xi_c$, the Higgs becomes unstable and evolves toward its nonzero VEV. The resulting waterfall transition simultaneously terminates inflation and triggers EWSB.

Our main finding is that the model can simultaneously accommodate constraints from cosmological and electroweak observations. The Higgs-sector parameters are fixed by electroweak-scale measurements at the LHC, while the inflationary parameters are constrained by CMB observables. We identify parameter regions that reproduce the measured Higgs properties while remaining compatible with the ACT DR6, Planck, and BAO constraints. {The large hierarchy between the inflationary and electroweak scales naturally suppresses $f_{\rm NL}$ to a negligible level.}

The essential ingredient that makes this construction viable is the interplay between gauged scale symmetry and the relaxation mechanism. After gauged scale symmetry is broken, a large effective vacuum-energy contribution $V_0$ is generated, providing the vacuum energy required to sustain inflation. To avoid leaving the Universe with such a large vacuum energy after inflation, we adopt the relaxation mechanism originally proposed to address the C. C. problem~\cite{Abbott:1984qf,Graham:2019bfu}. Since the underlying inflaton direction is axion-like, a small symmetry-breaking term $-k \theta_a$ can be introduced while leaving the slow-roll dynamics essentially unchanged. After the waterfall transition, the axion-like field can acquire a large displacement $v_a$, generating a contribution $-k v_a$ that compensates the inflationary vacuum energy. Although the resulting low-energy vacuum is metastable due to the tilted periodic potential, its lifetime can be much longer than the age of the Universe in the parameter region considered here. 

Several limitations of the present analysis should be pointed. First, the model is formulated as an effective theory, and the microscopic origin of the soft symmetry-breaking terms required by the relaxation mechanism remains to be clarified. Second, the connection between the inflationary and electroweak sectors has been studied mainly at tree level. Quantum corrections, RG evolution, and possible threshold effects may modify this connection and should be incorporated in a more complete analysis~\cite{Ren:2014sya,Barrie:2021mwi}. Finally, although the dependence on the reheating history has been parametrized through the number of $e$-folds, the postinflationary dynamics has not been studied explicitly.

Looking ahead, a detailed analysis of reheating is particularly important. The interactions between the axion-like scalar and the Higgs sector may lead to characteristic reheating channels and determine the reheating temperature, with possible implications for baryogenesis and dark matter production~\cite{Albrecht:1982wi,Kofman:1994rk,Kofman:1997yn,Krauss:1999ng,Garcia-Bellido:1999xos,Giudice:2000ex,Allahverdi:2010xz,Han:2024qbw}. Future CMB measurements~\cite{SimonsObservatory:2018koc,LiteBIRD:2022cnt} will further constrain the inflationary sector, while improved measurements of Higgs properties and searches for additional scalar states at future colliders~\cite{ATLAS:2025eii,FCC:2025lpp,Ai:2025cpj} can probe the associated low-energy phenomenology. These complementary probes will provide increasingly stringent tests of the proposed connection between inflationary dynamics and electroweak symmetry breaking.

\begin{acknowledgments} 
M.~Huang was supported in part by the National Natural Science Foundation of China (NSFC) under Grant Nos.~12235016 and 12221005, and by the Fundamental Research Funds for the Central Universities. Q.-S.~Yan was supported by the NSFC under Grant Nos.~11875260 and 12275143.
\end{acknowledgments} 

%
%%%%%%%%%%%%%%%%%%%%%%%%%%%%%%%

\appendix
\section{\label{d.o.f.}Analysis of the Degrees of Freedom}
Quadratic gravity, which is a renormalizable theory of gravity while also suffering from the ghost problem~\cite{Stelle:1976gc}, has a form similar to that of our scale-symmetric model. The quadratic-type action in our model, arising from the completeness of model construction based on gauged scale symmetry, serves the requirement for extending the defined nonminimal couplings $\beta$ and $\gamma$ in Eq.~\eqref{LagarangianUV} to $\chi_D'$ and $\chi_H'$ in Eq.~\eqref{eq:dilaton_Higgs}, which provides possible parameterized senarios for the following discussions. Since the reduction of quadratic gravity to linearized gravity and the associated ghost problem do not significantly affect our results, we do not discuss them further. The analysis of the degrees of freedom in our model is based on linearized gravity, which is sufficient for our purposes.

The degrees of freedom (d.o.f.) in scalar-tensor theories are subtle due to the presence of nonminimal coupling terms, requiring a detailed analysis of the kinetic matrix. We restrict the analysis to the scalar fields because the additional terms modify only the scalar sector. Based on Eq.~\eqref{LagarangianUV} and the definition of the BD field $\Theta$ in Eq.~\eqref{eq:dilaton_Higgs}, the kinetic terms of the scalar sector can be expressed as
\begin{equation}
\begin{aligned}
\sqrt{-g}\,\mathcal{L}_{\text{scalar}}
 &=\sqrt{-g}\,\Bigl[-\frac{1}{2}\Theta^2\hat R\\
 &\quad +(\hat\nabla_\mu\Phi)^*\hat\nabla^\mu\Phi
 +(\hat\nabla_\mu\phi)^\dagger\hat\nabla^\mu\phi\\
 &\quad +\frac{\delta_b}{2}
 (\hat\nabla_\mu\Phi\,\hat\nabla^\mu\Phi+\text{h.c.})\Bigr]\,.
\end{aligned}
\end{equation}
with the definition in Eq.~\eqref{case1}.

To separate dynamical and constrained variables, we introduce the induced metric
\begin{equation}
	h_{\mu \nu} = g_{\mu \nu} - n_\mu n_\nu\,.
\end{equation}
Then, the Ricci scalar is decomposed as
\begin{equation}
	R = {}^{(3)}R - K_{\mu \nu} K^{\mu \nu} + K^2 + 2 \nabla_\mu (a^\mu - n^\mu K)\,,
\end{equation}
where $K_{\mu \nu} \equiv h_\mu^\rho \nabla_\rho n_\nu$ is the extrinsic curvature, $K = K_\mu^\mu$ is the trace of the extrinsic curvature, and $a^\mu = n^\rho \nabla_\rho n^\mu$ is the acceleration. 
The presence of kinetic mixing between gravity and the scalar field motivates the ADM decomposition~\cite{Arnowitt:1962hi} of the metric
\begin{equation}
ds^2 = N^2 dt^2 - \gamma_{ij}\left(dx^i + N^i dt\right)\left(dx^j + N^j dt\right)\,,
\end{equation}
for the d.o.f. analysis. For compact notation, define
\begin{equation*}
\begin{aligned}
s_\theta&\equiv\sin\check\theta\,, &
c_\theta&\equiv\cos\check\theta\,,\\
C_a&\equiv\cos(2\check\theta_a)\,, &
S_a&\equiv\sin(2\check\theta_a)\,,\\
q&\equiv\frac{\chi_H^2}{\chi_D^2}\,.
\end{aligned}
\end{equation*}
Thus, the kinetic terms of the scalar sector can be expressed as
\begin{equation}
K=\begin{pmatrix}
-\dfrac{\Theta^2}{12} & \dfrac{\Theta}{2} & 0 & 0 \\[0.5em]
\dfrac{\Theta}{2} & K_{22} & K_{23} & K_{24} \\[0.5em]
0 & K_{23} & K_{33} & K_{34} \\[0.5em]
0 & K_{24} & K_{34} & K_{44}
\end{pmatrix}\,,
\end{equation}
where
\begin{align*}
K_{22}&=\frac{(1+\delta_b C_a)c_\theta^2}{\chi_D^2}
       +\frac{s_\theta^2}{\chi_H^2}\,,\\
K_{23}&=\Theta\Bigl(\frac{1}{\chi_H^2}
       -\frac{\delta_b C_a}{\chi_D^2}\Bigr)c_\theta s_\theta\,,\\
K_{24}&=-\frac{\delta_b S_a\Theta c_\theta^2}{\chi_D^2}\,,\\
K_{33}&=\frac{\Theta^2 c_\theta^2}{\chi_H^2}
       +\frac{\Theta^2(1+\delta_b C_a)s_\theta^2}{\chi_D^2}\,,\\
K_{34}&=\frac{\delta_b S_a\Theta^2c_\theta s_\theta}{\chi_D^2}\,,\\
K_{44}&=\frac{(1-\delta_b C_a)\Theta^2c_\theta^2}{\chi_D^2}\,.
\end{align*}
It turns out
\begin{equation}
\begin{aligned}
&\det K(\Theta,\check\theta,\check\theta_a)\\
&=\biggl\lvert\frac{\Theta^6 q}{12\chi_H^6}(1-s_\theta^2)
 \biggl\{q\Bigl[2q s_\theta^4\\
&\quad -(2q+3\chi_H^2)s_\theta^2-1\Bigr]\delta_b^2\\
&\quad -C_a c_\theta^2\Bigl[3\chi_H^2-s_\theta^2q(2+q)\Bigr]\delta_b\\
&\quad +q+3\chi_H^2
 -\Bigl[q(2-q)+3(1-q)\chi_H^2\Bigr]s_\theta^2\\
&\quad +(2-q)q s_\theta^4\biggr\}\biggr\rvert\,.
\end{aligned}
\label{eq:kineticmatrix}
\end{equation}
There are two representative cases, $\delta_b = 0$ and $\delta_b = 1$.
\begin{itemize}
	\item $\delta_b = 0$: This corresponds to the conventional case without a PQ symmetry breaking kinetic term. Eq~\eqref{eq:kineticmatrix} yields $\det K \neq 0$ identically, implying that the kinetic matrix is invertible and all scalar fields remain dynamical, consistent with the conventional case.
	\item $\delta_b = 1$: This corresponds to the scenario with maximal PQ symmetry breaking in the kinetic sector, adopted in our model. The kinetic matrix is simplified to
	\begin{equation}
\begin{aligned}
\det K &\propto\biggl\lvert(1-\cos^2\check\theta_a)(1-s_\theta^2)\\
 &\qquad\times\Bigl[(2+q)q\frac{s_\theta^2}{\chi_H^2}-3\Bigr]
 \biggr\rvert\\
 &\propto\biggl\lvert3-(2+q)q\frac{U^2}{f^2}\biggr\rvert\,,
\end{aligned}
\end{equation}
	where the vevs of the scalar fields are substituted in the second line. The kinetic matrix is invertible unless
	\begin{equation}
		\frac{\chi_H^2}{\chi_D^2} =\sqrt{1 + \frac{3 f^2}{U^2}} - 1\sim \sqrt{1 + \frac{3 M_P^2}{v_{sm}^2}} - 1
		\,.
	\end{equation}
\end{itemize}

Therefore, the PQ symmetry breaking kinetic term is a viable extension of the theory, with the kinetic matrix remaining invertible over a broad parameter range and the d.o.f. structure unaffected.
\bibliography{references}

@article{Lyth:1998xn,
    author = "Lyth, David H. and Riotto, Antonio",
    title = "{Particle physics models of inflation and the cosmological density perturbation}",
    eprint = "hep-ph/9807278",
    archivePrefix = "arXiv",
    reportNumber = "LANCS-TH-9720, FERMILAB-PUB-97-292-A, CERN-TH-97-383, OUTP-98-39-P",
    doi = "10.1016/S0370-1573(98)00128-8",
    journal = "Phys. Rept.",
    volume = "314",
    pages = "1--146",
    year = "1999"
}

@article{Dvali:1994ms,
    author = "Dvali, G. R. and Shafi, Q. and Schaefer, Robert K.",
    title = "{Large scale structure and supersymmetric inflation without fine tuning}",
    eprint = "hep-ph/9406319",
    archivePrefix = "arXiv",
    reportNumber = "BA-94-32",
    doi = "10.1103/PhysRevLett.73.1886",
    journal = "Phys. Rev. Lett.",
    volume = "73",
    pages = "1886--1889",
    year = "1994"
}

@article{Ballesteros:2016euj,
    author = "Ballesteros, Guillermo and Redondo, Javier and Ringwald, Andreas and Tamarit, Carlos",
    title = "{Unifying inflation with the axion, dark matter, baryogenesis and the seesaw mechanism}",
    eprint = "1608.05414",
    archivePrefix = "arXiv",
    primaryClass = "hep-ph",
    reportNumber = "DESY-16-049, IPPP-16-25",
    doi = "10.1103/PhysRevLett.118.071802",
    journal = "Phys. Rev. Lett.",
    volume = "118",
    number = "7",
    pages = "071802",
    year = "2017"
}

@article{Mukhanov:1981xt,
    author = "Mukhanov, Viatcheslav F. and Chibisov, G. V.",
    title = "{Quantum Fluctuations and a Nonsingular Universe}",
    journal = "JETP Lett.",
    volume = "33",
    pages = "532--535",
    year = "1981"
}

@article{Guth:1982ec,
    author = "Guth, Alan H. and Pi, S. Y.",
    title = "{Fluctuations in the New Inflationary Universe}",
    doi = "10.1103/PhysRevLett.49.1110",
    journal = "Phys. Rev. Lett.",
    volume = "49",
    pages = "1110--1113",
    year = "1982"
}

@article{Planck:2018jri,
    author = "Akrami, Y. and others",
    collaboration = "Planck",
    title = "{Planck 2018 results. X. Constraints on inflation}",
    eprint = "1807.06211",
    archivePrefix = "arXiv",
    primaryClass = "astro-ph.CO",
    doi = "10.1051/0004-6361/201833887",
    journal = "Astron. Astrophys.",
    volume = "641",
    pages = "A10",
    year = "2020"
}

@article{Guth:1980zm,
    author = "Guth, Alan H.",
    editor = "Fang, Li-Zhi and Ruffini, R.",
    title = "{The Inflationary Universe: A Possible Solution to the Horizon and Flatness Problems}",
    reportNumber = "SLAC-PUB-2576",
    doi = "10.1103/PhysRevD.23.347",
    journal = "Phys. Rev. D",
    volume = "23",
    pages = "347--356",
    year = "1981"
}

@article{Linde:1981mu,
    author = "Linde, Andrei D.",
    editor = "Fang, Li-Zhi and Ruffini, R.",
    title = "{A New Inflationary Universe Scenario: A Possible Solution of the Horizon, Flatness, Homogeneity, Isotropy and Primordial Monopole Problems}",
    reportNumber = "LEBEDEV-81-229",
    doi = "10.1016/0370-2693(82)91219-9",
    journal = "Phys. Lett. B",
    volume = "108",
    pages = "389--393",
    year = "1982"
}

@article{Albrecht:1982wi,
    author = "Albrecht, Andreas and Steinhardt, Paul J.",
    editor = "Fang, Li-Zhi and Ruffini, R.",
    title = "{Cosmology for Grand Unified Theories with Radiatively Induced Symmetry Breaking}",
    reportNumber = "UPR-0185T",
    doi = "10.1103/PhysRevLett.48.1220",
    journal = "Phys. Rev. Lett.",
    volume = "48",
    pages = "1220--1223",
    year = "1982"
}

@article{Yin:2025rrs,
    author = "Yin, Wen",
    title = "{Higgs-like inflation ACTivated mass}",
    eprint = "2505.03004",
    archivePrefix = "arXiv",
    primaryClass = "hep-ph",
    doi = "10.1088/1475-7516/2025/09/062",
    journal = "JCAP",
    volume = "09",
    pages = "062",
    year = "2025"
}

@article{Kallosh:2026qrc,
    author = "Kallosh, Renata and Linde, Andrei",
    title = "{New exponential and polynomial {\ensuremath{\xi}}-attractors}",
    eprint = "2605.04415",
    archivePrefix = "arXiv",
    primaryClass = "hep-th",
    doi = "10.1088/1475-7516/2026/09/027",
    journal = "JCAP",
    volume = "09",
    pages = "027",
    year = "2026"
}

@article{Yuennan:2025inm,
    author = "Yuennan, Jureeporn and Atamurotov, Farruh and Channuie, Phongpichit",
    title = "{Radiative-corrected Higgs inflation in light of the latest ACT observations}",
    eprint = "2510.05770",
    archivePrefix = "arXiv",
    primaryClass = "astro-ph.CO",
    doi = "10.1016/j.physletb.2025.139958",
    journal = "Phys. Lett. B",
    volume = "871",
    pages = "139958",
    year = "2025"
}

@article{Ferreira:2025lrd,
    author = "Ferreira, Elisa G. M. and McDonough, Evan and Balkenhol, Lennart and Kallosh, Renata and Knox, Lloyd and Linde, Andrei",
    title = "{BAO-CMB tension and implications for inflation}",
    eprint = "2507.12459",
    archivePrefix = "arXiv",
    primaryClass = "astro-ph.CO",
    doi = "10.1103/lq71-b84v",
    journal = "Phys. Rev. D",
    volume = "113",
    number = "4",
    pages = "043524",
    year = "2026"
}

@article{McDonough:2025lzo,
    author = "McDonough, Evan and Ferreira, Elisa G. M.",
    title = "{Constraints on primordial power spectrum parameters from DESI and CMB data}",
    eprint = "2512.05108",
    archivePrefix = "arXiv",
    primaryClass = "astro-ph.CO",
    doi = "10.1103/9fjv-tl3l",
    journal = "Phys. Rev. D",
    volume = "114",
    number = "2",
    pages = "023559",
    year = "2026"
}

@article{Balkenhol:2025wms,
    author = "Balkenhol, L. and others",
    title = "{Inflation at the End of 2025: Constraints on $r$ and $n_s$ Using the Latest CMB and BAO Data}",
    eprint = "2512.10613",
    archivePrefix = "arXiv",
    primaryClass = "astro-ph.CO",
    doi = "10.33232/001c.164435",
    month = "12",
    year = "2025"
}

@article{Ishida:2019wkd,
    author = "Ishida, Hiroyuki and Matsuzaki, Shinya",
    title = "{A Walking Dilaton Inflation}",
    eprint = "1912.09740",
    archivePrefix = "arXiv",
    primaryClass = "hep-ph",
    reportNumber = "KEK-TH-2181",
    doi = "10.1016/j.physletb.2020.135390",
    journal = "Phys. Lett. B",
    volume = "804",
    pages = "135390",
    year = "2020"
}

@article{Zhang:2023acu,
    author = "Zhang, He-Xu and Matsuzaki, Shinya and Ishida, Hiroyuki",
    title = "{Dynamical realization of the small field inflation of Coleman-Weinberg type in the post supercooled universe}",
    eprint = "2306.15471",
    archivePrefix = "arXiv",
    primaryClass = "hep-ph",
    doi = "10.1016/j.physletb.2023.138256",
    journal = "Phys. Lett. B",
    volume = "846",
    pages = "138256",
    year = "2023"
}

@article{Garcia-Bellido:2008ycs,
    author = "Garcia-Bellido, Juan and Figueroa, Daniel G. and Rubio, Javier",
    title = "{Preheating in the Standard Model with the Higgs-Inflaton coupled to gravity}",
    eprint = "0812.4624",
    archivePrefix = "arXiv",
    primaryClass = "hep-ph",
    reportNumber = "IFT-UAM-CSIC-08-93",
    doi = "10.1103/PhysRevD.79.063531",
    journal = "Phys. Rev. D",
    volume = "79",
    pages = "063531",
    year = "2009"
}

@article{Rubio:2018ogq,
    author = "Rubio, Javier",
    title = "{Higgs inflation}",
    eprint = "1807.02376",
    archivePrefix = "arXiv",
    primaryClass = "hep-ph",
    doi = "10.3389/fspas.2018.00050",
    journal = "Front. Astron. Space Sci.",
    volume = "5",
    pages = "50",
    year = "2019"
}

@article{Appelquist:2022mjb,
    author = "Appelquist, Thomas and Ingoldby, James and Piai, Maurizio",
    title = "{Dilaton Effective Field Theory}",
    eprint = "2209.14867",
    archivePrefix = "arXiv",
    primaryClass = "hep-ph",
    doi = "10.3390/universe9010010",
    journal = "Universe",
    volume = "9",
    number = "1",
    pages = "10",
    year = "2023"
}

@article{Migdal:1982jp,
    author = "Migdal, Alexander A. and Shifman, Mikhail A.",
    title = "{Dilaton Effective Lagrangian in Gluodynamics}",
    reportNumber = "ITEP-15-1982",
    doi = "10.1016/0370-2693(82)90089-2",
    journal = "Phys. Lett. B",
    volume = "114",
    pages = "445--449",
    year = "1982"
}

@article{Cata:2018wzl,
    author = "Cat{\`a}, O. and Crewther, R. J. and Tunstall, Lewis C.",
    title = "{Crawling technicolor}",
    eprint = "1803.08513",
    archivePrefix = "arXiv",
    primaryClass = "hep-ph",
    reportNumber = "SI-HEP-2018-09, QFET-2018-05, ADP-18-3/T1051, ADP-18-3-T1051",
    doi = "10.1103/PhysRevD.100.095007",
    journal = "Phys. Rev. D",
    volume = "100",
    number = "9",
    pages = "095007",
    year = "2019"
}

@article{Zwicky:2023krx,
    author = "Zwicky, Roman",
    title = "{QCD with an infrared fixed point and a dilaton}",
    eprint = "2312.13761",
    archivePrefix = "arXiv",
    primaryClass = "hep-ph",
    reportNumber = "CERN-TH-2023-201",
    doi = "10.1103/PhysRevD.110.014048",
    journal = "Phys. Rev. D",
    volume = "110",
    number = "1",
    pages = "014048",
    year = "2024",
    note = "[Erratum: Phys.Rev.D 113, 039902 (2026)]"
}

@article{Jarv:2016sow,
    author = {J{\"a}rv, Laur and Kannike, Kristjan and Marzola, Luca and Racioppi, Antonio and Raidal, Martti and R{\"u}nkla, Mihkel and Saal, Margus and Veerm{\"a}e, Hardi},
    title = "{Frame-Independent Classification of Single-Field Inflationary Models}",
    eprint = "1612.06863",
    archivePrefix = "arXiv",
    primaryClass = "hep-ph",
    doi = "10.1103/PhysRevLett.118.151302",
    journal = "Phys. Rev. Lett.",
    volume = "118",
    number = "15",
    pages = "151302",
    year = "2017"
}

@article{McAllister:2008hb,
    author = "McAllister, Liam and Silverstein, Eva and Westphal, Alexander",
    title = "{Gravity Waves and Linear Inflation from Axion Monodromy}",
    eprint = "0808.0706",
    archivePrefix = "arXiv",
    primaryClass = "hep-th",
    reportNumber = "SLAC-PUB-13357, SU-ITP-08-15",
    doi = "10.1103/PhysRevD.82.046003",
    journal = "Phys. Rev. D",
    volume = "82",
    pages = "046003",
    year = "2010"
}

@article{McAllister:2023vgy,
    author = "McAllister, Liam and Quevedo, Fernando",
    title = "{Moduli Stabilization in String Theory}",
    eprint = "2310.20559",
    archivePrefix = "arXiv",
    primaryClass = "hep-th",
    month = "10",
    year = "2023"
}

@article{Kallosh:2002gf,
    author = "Kallosh, Renata and Linde, Andrei D. and Prokushkin, Sergey and Shmakova, Marina",
    title = "{Supergravity, dark energy and the fate of the universe}",
    eprint = "hep-th/0208156",
    archivePrefix = "arXiv",
    reportNumber = "SLAC-PUB-9408",
    doi = "10.1103/PhysRevD.66.123503",
    journal = "Phys. Rev. D",
    volume = "66",
    pages = "123503",
    year = "2002"
}

@article{Planck:2018vyg,
    author = "Aghanim, N. and others",
    collaboration = "Planck",
    title = "{Planck 2018 results. VI. Cosmological parameters}",
    eprint = "1807.06209",
    archivePrefix = "arXiv",
    primaryClass = "astro-ph.CO",
    doi = "10.1051/0004-6361/201833910",
    journal = "Astron. Astrophys.",
    volume = "641",
    pages = "A6",
    year = "2020",
    note = "[Erratum: Astron.Astrophys. 652, C4 (2021)]"
}

@article{AtacamaCosmologyTelescope:2025blo,
    author = "Louis, Thibaut and others",
    collaboration = "Atacama Cosmology Telescope",
    title = "{The Atacama Cosmology Telescope: DR6 power spectra, likelihoods and {\ensuremath{\Lambda}}CDM parameters}",
    eprint = "2503.14452",
    archivePrefix = "arXiv",
    primaryClass = "astro-ph.CO",
    reportNumber = "FERMILAB-PUB-25-0071-PPD",
    doi = "10.1088/1475-7516/2025/11/062",
    journal = "JCAP",
    volume = "11",
    pages = "062",
    year = "2025"
}

@article{AtacamaCosmologyTelescope:2025nti,
    author = "Calabrese, Erminia and others",
    collaboration = "Atacama Cosmology Telescope",
    title = "{The Atacama Cosmology Telescope: DR6 constraints on extended cosmological models}",
    eprint = "2503.14454",
    archivePrefix = "arXiv",
    primaryClass = "astro-ph.CO",
    reportNumber = "FERMILAB-PUB-25-0157-PPD",
    doi = "10.1088/1475-7516/2025/11/063",
    journal = "JCAP",
    volume = "11",
    pages = "063",
    year = "2025"
}

@article{Stelle:1976gc,
    author = "Stelle, K. S.",
    title = "{Renormalization of Higher Derivative Quantum Gravity}",
    reportNumber = "PRINT-76-1059 (BRANDEIS)",
    doi = "10.1103/PhysRevD.16.953",
    journal = "Phys. Rev. D",
    volume = "16",
    pages = "953--969",
    year = "1977"
}

@article{Linde:1983gd,
    author = "Linde, Andrei D.",
    title = "{Chaotic Inflation}",
    doi = "10.1016/0370-2693(83)90837-7",
    journal = "Phys. Lett. B",
    volume = "129",
    pages = "177--181",
    year = "1983"
}

@article{Bezrukov:2007ep,
    author = "Bezrukov, Fedor L. and Shaposhnikov, Mikhail",
    title = "{The Standard Model Higgs boson as the inflaton}",
    eprint = "0710.3755",
    archivePrefix = "arXiv",
    primaryClass = "hep-th",
    doi = "10.1016/j.physletb.2007.11.072",
    journal = "Phys. Lett. B",
    volume = "659",
    pages = "703--706",
    year = "2008"
}

@article{Bezrukov:2010jz,
    author = "Bezrukov, F. and Magnin, A. and Shaposhnikov, M. and Sibiryakov, S.",
    title = "{Higgs inflation: consistency and generalisations}",
    eprint = "1008.5157",
    archivePrefix = "arXiv",
    primaryClass = "hep-ph",
    doi = "10.1007/JHEP01(2011)016",
    journal = "JHEP",
    volume = "01",
    pages = "016",
    year = "2011"
}

@article{Burgess:2009ea,
    author = "Burgess, C. P. and Lee, Hyun Min and Trott, Michael",
    title = "{Power-counting and the Validity of the Classical Approximation During Inflation}",
    eprint = "0902.4465",
    archivePrefix = "arXiv",
    primaryClass = "hep-ph",
    reportNumber = "PI-PARTPHYS-121",
    doi = "10.1088/1126-6708/2009/09/103",
    journal = "JHEP",
    volume = "09",
    pages = "103",
    year = "2009"
}

@article{Barbon:2009ya,
    author = "Barbon, J. L. F. and Espinosa, J. R.",
    title = "{On the Naturalness of Higgs Inflation}",
    eprint = "0903.0355",
    archivePrefix = "arXiv",
    primaryClass = "hep-ph",
    reportNumber = "IFT-UAM-CSIC-09-10, UAB-FT-665",
    doi = "10.1103/PhysRevD.79.081302",
    journal = "Phys. Rev. D",
    volume = "79",
    pages = "081302",
    year = "2009"
}

@article{Burgess:2010zq,
    author = "Burgess, C. P. and Lee, Hyun Min and Trott, Michael",
    title = "{Comment on Higgs Inflation and Naturalness}",
    eprint = "1002.2730",
    archivePrefix = "arXiv",
    primaryClass = "hep-ph",
    reportNumber = "CERN-PH-TH-2010-033, PI-PARTPHYS-174, CERN---PH---TH--2010-033",
    doi = "10.1007/JHEP07(2010)007",
    journal = "JHEP",
    volume = "07",
    pages = "007",
    year = "2010"
}

@article{Starobinsky:1980te,
    author = "Starobinsky, Alexei A.",
    editor = "Khalatnikov, I. M. and Mineev, V. P.",
    title = "{A New Type of Isotropic Cosmological Models Without Singularity}",
    doi = "10.1016/0370-2693(80)90670-X",
    journal = "Phys. Lett. B",
    volume = "91",
    pages = "99--102",
    year = "1980"
}

@article{Ghilencea:2018thl,
    author = "Ghilencea, D. M. and Lee, Hyun Min",
    title = "{Weyl gauge symmetry and its spontaneous breaking in the standard model and inflation}",
    eprint = "1809.09174",
    archivePrefix = "arXiv",
    primaryClass = "hep-th",
    doi = "10.1103/PhysRevD.99.115007",
    journal = "Phys. Rev. D",
    volume = "99",
    number = "11",
    pages = "115007",
    year = "2019"
}

@article{deCesare:2016mml,
    author = "de Cesare, Marco and Moffat, John W. and Sakellariadou, Mairi",
    title = "{Local conformal symmetry in non-Riemannian geometry and the origin of physical scales}",
    eprint = "1612.08066",
    archivePrefix = "arXiv",
    primaryClass = "hep-th",
    reportNumber = "KCL-PH-TH-2016-71",
    doi = "10.1140/epjc/s10052-017-5183-0",
    journal = "Eur. Phys. J. C",
    volume = "77",
    number = "9",
    pages = "605",
    year = "2017"
}

@article{Aoki:2022csb,
    author = "Aoki, Shuntaro and Lee, Hyun Min",
    title = "{Weyl gravity extension of Higgs inflation}",
    eprint = "2207.05484",
    archivePrefix = "arXiv",
    primaryClass = "hep-ph",
    doi = "10.1103/PhysRevD.108.035045",
    journal = "Phys. Rev. D",
    volume = "108",
    number = "3",
    pages = "035045",
    year = "2023"
}

@article{Linde:1993cn,
    author = "Linde, Andrei D.",
    title = "{Hybrid inflation}",
    eprint = "astro-ph/9307002",
    archivePrefix = "arXiv",
    reportNumber = "SU-ITP-93-17",
    doi = "10.1103/PhysRevD.49.748",
    journal = "Phys. Rev. D",
    volume = "49",
    pages = "748--754",
    year = "1994"
}

@article{Kallosh:2013hoa,
    author = "Kallosh, Renata and Linde, Andrei",
    title = "{Universality Class in Conformal Inflation}",
    eprint = "1306.5220",
    archivePrefix = "arXiv",
    primaryClass = "hep-th",
    doi = "10.1088/1475-7516/2013/07/002",
    journal = "JCAP",
    volume = "07",
    pages = "002",
    year = "2013"
}

@article{Abbott:1984qf,
    author = "Abbott, L. F.",
    title = "{A Mechanism for Reducing the Value of the Cosmological Constant}",
    reportNumber = "BRX-TH-175",
    doi = "10.1016/0370-2693(85)90459-9",
    journal = "Phys. Lett. B",
    volume = "150",
    pages = "427--430",
    year = "1985"
}

@article{Graham:2015cka,
    author = "Graham, Peter W. and Kaplan, David E. and Rajendran, Surjeet",
    title = "{Cosmological Relaxation of the Electroweak Scale}",
    eprint = "1504.07551",
    archivePrefix = "arXiv",
    primaryClass = "hep-ph",
    doi = "10.1103/PhysRevLett.115.221801",
    journal = "Phys. Rev. Lett.",
    volume = "115",
    number = "22",
    pages = "221801",
    year = "2015"
}

@article{Graham:2019bfu,
    author = "Graham, Peter W. and Kaplan, David E. and Rajendran, Surjeet",
    title = "{Relaxation of the Cosmological Constant}",
    eprint = "1902.06793",
    archivePrefix = "arXiv",
    primaryClass = "hep-ph",
    doi = "10.1103/PhysRevD.100.015048",
    journal = "Phys. Rev. D",
    volume = "100",
    number = "1",
    pages = "015048",
    year = "2019"
}

@article{Ghilencea:2021lpa,
    author = "Ghilencea, D. M.",
    title = "{Standard Model in Weyl conformal geometry}",
    eprint = "2104.15118",
    archivePrefix = "arXiv",
    primaryClass = "hep-ph",
    doi = "10.1140/epjc/s10052-021-09887-y",
    journal = "Eur. Phys. J. C",
    volume = "82",
    number = "1",
    pages = "23",
    year = "2022"
}

@article{Wu:2025hfp,
    author = "Wu, J. E. and Yan, Q. S.",
    title = "{Confronting a dilaton model with LHC measurements}",
    eprint = "2501.07820",
    archivePrefix = "arXiv",
    primaryClass = "hep-ph",
    doi = "10.1103/ycxz-bvfn",
    journal = "Phys. Rev. D",
    volume = "113",
    number = "5",
    pages = "055049",
    year = "2026"
}

@book{david_h_lyth_primordial_nodate,
	title = {The primordial density perturbation cosmology, inflation and the origin of structure},
	author = {{David H. Lyth} and {Andrew R. Liddle}}
}

@article{Stewart:1993bc,
    author = "Stewart, Ewan D. and Lyth, David H.",
    title = "{A More accurate analytic calculation of the spectrum of cosmological perturbations produced during inflation}",
    eprint = "gr-qc/9302019",
    archivePrefix = "arXiv",
    reportNumber = "KUNS-1176, LANCS-TH-93-01",
    doi = "10.1016/0370-2693(93)90379-V",
    journal = "Phys. Lett. B",
    volume = "302",
    pages = "171--175",
    year = "1993"
}

@article{Dai:2014jja,
    author = "Dai, Liang and Kamionkowski, Marc and Wang, Junpu",
    title = "{Reheating constraints to inflationary models}",
    eprint = "1404.6704",
    archivePrefix = "arXiv",
    primaryClass = "astro-ph.CO",
    doi = "10.1103/PhysRevLett.113.041302",
    journal = "Phys. Rev. Lett.",
    volume = "113",
    pages = "041302",
    year = "2014"
}

@article{Cook:2015vqa,
    author = "Cook, Jessica L. and Dimastrogiovanni, Emanuela and Easson, Damien A. and Krauss, Lawrence M.",
    title = "{Reheating predictions in single field inflation}",
    eprint = "1502.04673",
    archivePrefix = "arXiv",
    primaryClass = "astro-ph.CO",
    doi = "10.1088/1475-7516/2015/04/047",
    journal = "JCAP",
    volume = "04",
    pages = "047",
    year = "2015"
}

@article{Coleman:1977py,
    author = "Coleman, Sidney R.",
    title = "{The Fate of the False Vacuum. 1. Semiclassical Theory}",
    reportNumber = "HUTP-77-A004",
    doi = "10.1103/PhysRevD.16.1248",
    journal = "Phys. Rev. D",
    volume = "15",
    pages = "2929--2936",
    year = "1977",
    note = "[Erratum: Phys.Rev.D 16, 1248 (1977)]"
}

@article{Gordon:2000hv,
    author = "Gordon, Christopher and Wands, David and Bassett, Bruce A. and Maartens, Roy",
    title = "{Adiabatic and entropy perturbations from inflation}",
    eprint = "astro-ph/0009131",
    archivePrefix = "arXiv",
    doi = "10.1103/PhysRevD.63.023506",
    journal = "Phys. Rev. D",
    volume = "63",
    pages = "023506",
    year = "2000"
}

@article{peccei_cp_1977,
	title = {{CP} Conservation in the Presence of Instantons},
	volume = {38},
	doi = {10.1103/PhysRevLett.38.1440},
	pages = {1440--1443},
	journaltitle = {Phys. Rev. Lett.},
	author = {Peccei, R. D. and Quinn, Helen R.},
	date = {1977},
}

@article{hehl_metric_1995,
   title={Metric-affine gauge theory of gravity: field equations, Noether identities, world spinors, and breaking of dilation invariance},
   volume={258},
   ISSN={0370-1573},
   url={http://dx.doi.org/10.1016/0370-1573(94)00111-F},
   DOI={10.1016/0370-1573(94)00111-f},
   number={1–2},
   journal={Physics Reports},
   publisher={Elsevier BV},
   author={Hehl, Friedrich W. and McCrea, J.Dermott and Mielke, Eckehard W. and Ne’eman, Yuval},
   year={1995},
   month=jul, pages={1–171} }

@article{Dirac:1973gk,
    author = "Dirac, Paul A. M.",
    title = "{Long range forces and broken symmetries}",
    doi = "10.1098/rspa.1973.0070",
    journal = "Proc. Roy. Soc. Lond. A",
    volume = "333",
    pages = "403--418",
    year = "1973"
}

@article{trautman_geometry_1979,
    author = "Trautman, A.",
    title = "{THE GEOMETRY OF GAUGE FIELDS. (TALK)}",
    doi = "10.1007/BF01603811",
    journal = "Czech. J. Phys. B",
    volume = "29",
    pages = "107--116",
    year = "1979"
}

@article{Flato:1987bb,
    author = "Flato, Moshe and Raczka, Ryszard",
    title = "{A Possible Gravitational Origin of Higgs Field in the Standard Model}",
    reportNumber = "SISSA-107-87-EP",
    doi = "10.1016/0370-2693(88)91213-0",
    journal = "Phys. Lett. B",
    volume = "208",
    pages = "110--114",
    year = "1988"
}

@article{Capozziello:2011et,
    author = "Capozziello, Salvatore and De Laurentis, Mariafelicia",
    title = "{Extended Theories of Gravity}",
    eprint = "1108.6266",
    archivePrefix = "arXiv",
    primaryClass = "gr-qc",
    doi = "10.1016/j.physrep.2011.09.003",
    journal = "Phys. Rept.",
    volume = "509",
    pages = "167--321",
    year = "2011"
}

@article{SupernovaSearchTeam:1998fmf,
    author = "Riess, Adam G. and others",
    collaboration = "Supernova Search Team",
    title = "{Observational evidence from supernovae for an accelerating universe and a cosmological constant}",
    eprint = "astro-ph/9805201",
    archivePrefix = "arXiv",
    doi = "10.1086/300499",
    journal = "Astron. J.",
    volume = "116",
    pages = "1009--1038",
    year = "1998"
}

@article{SupernovaCosmologyProject:1998vns,
    author = "Perlmutter, S. and others",
    collaboration = "Supernova Cosmology Project",
    title = "{Measurements of $\Omega$ and $\Lambda$ from 42 High Redshift Supernovae}",
    eprint = "astro-ph/9812133",
    archivePrefix = "arXiv",
    reportNumber = "LBNL-41801, LBL-41801",
    doi = "10.1086/307221",
    journal = "Astrophys. J.",
    volume = "517",
    pages = "565--586",
    year = "1999"
}

@article{SimonsObservatory:2018koc,
    author = "Ade, Peter and others",
    collaboration = "Simons Observatory",
    title = "{The Simons Observatory: Science goals and forecasts}",
    eprint = "1808.07445",
    archivePrefix = "arXiv",
    primaryClass = "astro-ph.CO",
    doi = "10.1088/1475-7516/2019/02/056",
    journal = "JCAP",
    volume = "02",
    pages = "056",
    year = "2019"
}

@article{LiteBIRD:2022cnt,
    author = "Allys, E. and others",
    collaboration = "LiteBIRD",
    title = "{Probing Cosmic Inflation with the LiteBIRD Cosmic Microwave Background Polarization Survey}",
    eprint = "2202.02773",
    archivePrefix = "arXiv",
    primaryClass = "astro-ph.IM",
    doi = "10.1093/ptep/ptac150",
    journal = "PTEP",
    volume = "2023",
    number = "4",
    pages = "042F01",
    year = "2023"
}

@article{Kofman:1994rk,
    author = "Kofman, Lev and Linde, Andrei D. and Starobinsky, Alexei A.",
    title = "{Reheating after inflation}",
    eprint = "hep-th/9405187",
    archivePrefix = "arXiv",
    reportNumber = "UH-IFA-94-35, SU-ITP-94-13, YITP-U-94-15",
    doi = "10.1103/PhysRevLett.73.3195",
    journal = "Phys. Rev. Lett.",
    volume = "73",
    pages = "3195--3198",
    year = "1994"
}

@article{Kofman:1997yn,
    author = "Kofman, Lev and Linde, Andrei D. and Starobinsky, Alexei A.",
    title = "{Towards the theory of reheating after inflation}",
    eprint = "hep-ph/9704452",
    archivePrefix = "arXiv",
    reportNumber = "IFA-97-28, SU-ITP-97-18",
    doi = "10.1103/PhysRevD.56.3258",
    journal = "Phys. Rev. D",
    volume = "56",
    pages = "3258--3295",
    year = "1997"
}

@article{Giudice:2000ex,
    author = "Giudice, Gian Francesco and Kolb, Edward W. and Riotto, Antonio",
    title = "{Largest temperature of the radiation era and its cosmological implications}",
    eprint = "hep-ph/0005123",
    archivePrefix = "arXiv",
    reportNumber = "SNS-PH-00-05, FERMILAB-PUB-00-075-A, CERN-TH-2000-107",
    doi = "10.1103/PhysRevD.64.023508",
    journal = "Phys. Rev. D",
    volume = "64",
    pages = "023508",
    year = "2001"
}

@article{Allahverdi:2010xz,
    author = "Allahverdi, Rouzbeh and Brandenberger, Robert and Cyr-Racine, Francis-Yan and Mazumdar, Anupam",
    title = "{Reheating in Inflationary Cosmology: Theory and Applications}",
    eprint = "1001.2600",
    archivePrefix = "arXiv",
    primaryClass = "hep-th",
    doi = "10.1146/annurev.nucl.012809.104511",
    journal = "Ann. Rev. Nucl. Part. Sci.",
    volume = "60",
    pages = "27--51",
    year = "2010"
}

@article{Burgess:2014lza,
    author = "Burgess, C. P. and Patil, Subodh P. and Trott, Michael",
    title = "{On the Predictiveness of Single-Field Inflationary Models}",
    eprint = "1402.1476",
    archivePrefix = "arXiv",
    primaryClass = "hep-ph",
    reportNumber = "CERN-PH-TH-2014-024",
    doi = "10.1007/JHEP06(2014)010",
    journal = "JHEP",
    volume = "06",
    pages = "010",
    year = "2014"
}

@article{DeSimone:2008ei,
    author = "De Simone, Andrea and Hertzberg, Mark P. and Wilczek, Frank",
    title = "{Running Inflation in the Standard Model}",
    eprint = "0812.4946",
    archivePrefix = "arXiv",
    primaryClass = "hep-ph",
    reportNumber = "MIT-CTP-4008",
    doi = "10.1016/j.physletb.2009.05.054",
    journal = "Phys. Lett. B",
    volume = "678",
    pages = "1--8",
    year = "2009"
}

@article{Bezrukov:2008ej,
    author = "Bezrukov, Fedor L. and Magnin, Amaury and Shaposhnikov, Mikhail",
    title = "{Standard Model Higgs boson mass from inflation}",
    eprint = "0812.4950",
    archivePrefix = "arXiv",
    primaryClass = "hep-ph",
    doi = "10.1016/j.physletb.2009.03.035",
    journal = "Phys. Lett. B",
    volume = "675",
    pages = "88--92",
    year = "2009"
}

@article{Bezrukov:2009db,
    author = "Bezrukov, F. and Shaposhnikov, M.",
    title = "{Standard Model Higgs boson mass from inflation: Two loop analysis}",
    eprint = "0904.1537",
    archivePrefix = "arXiv",
    primaryClass = "hep-ph",
    doi = "10.1088/1126-6708/2009/07/089",
    journal = "JHEP",
    volume = "07",
    pages = "089",
    year = "2009"
}

@article{Allison:2013uaa,
    author = "Allison, Kyle",
    title = "{Higgs xi-inflation for the 125-126 GeV Higgs: a two-loop analysis}",
    eprint = "1306.6931",
    archivePrefix = "arXiv",
    primaryClass = "hep-ph",
    doi = "10.1007/JHEP02(2014)040",
    journal = "JHEP",
    volume = "02",
    pages = "040",
    year = "2014"
}

@article{Enckell:2016xse,
    author = "Enckell, Vera-Maria and Enqvist, Kari and Nurmi, Sami",
    title = "{Observational signatures of Higgs inflation}",
    eprint = "1603.07572",
    archivePrefix = "arXiv",
    primaryClass = "astro-ph.CO",
    doi = "10.1088/1475-7516/2016/07/047",
    journal = "JCAP",
    volume = "07",
    pages = "047",
    year = "2016"
}

@article{Ren:2014sya,
    author = "Ren, Jing and Xianyu, Zhong-Zhi and He, Hong-Jian",
    title = "{Higgs Gravitational Interaction, Weak Boson Scattering, and Higgs Inflation in Jordan and Einstein Frames}",
    eprint = "1404.4627",
    archivePrefix = "arXiv",
    primaryClass = "gr-qc",
    reportNumber = "KCL-PH-TH-2014-25",
    doi = "10.1088/1475-7516/2014/06/032",
    journal = "JCAP",
    volume = "06",
    pages = "032",
    year = "2014"
}

@article{Han:2024qbw,
    author = "Han, Chengcheng and He, Hong-Jian and Song, Linghao and You, Jingtao",
    title = "{Cosmological signatures of neutrino seesaw mechanism}",
    eprint = "2412.21045",
    archivePrefix = "arXiv",
    primaryClass = "hep-ph",
    doi = "10.1103/b7hv-3h2p",
    journal = "Phys. Rev. D",
    volume = "112",
    number = "8",
    pages = "L081309",
    year = "2025"
}

@article{Barrie:2021mwi,
    author = "Barrie, Neil D. and Han, Chengcheng and Murayama, Hitoshi",
    title = "{Affleck-Dine Leptogenesis from Higgs Inflation}",
    eprint = "2106.03381",
    archivePrefix = "arXiv",
    primaryClass = "hep-ph",
    doi = "10.1103/PhysRevLett.128.141801",
    journal = "Phys. Rev. Lett.",
    volume = "128",
    number = "14",
    pages = "141801",
    year = "2022"
}

@article{Sasaki:1995aw,
    author = "Sasaki, Misao and Stewart, Ewan D.",
    title = "{A General analytic formula for the spectral index of the density perturbations produced during inflation}",
    eprint = "astro-ph/9507001",
    archivePrefix = "arXiv",
    reportNumber = "LANCS-TH-9504, OU-TAP-22",
    doi = "10.1143/PTP.95.71",
    journal = "Prog. Theor. Phys.",
    volume = "95",
    pages = "71--78",
    year = "1996"
}

@article{Planck:2019kim,
    author = "Akrami, Y. and others",
    collaboration = "Planck",
    title = "{Planck 2018 results. IX. Constraints on primordial non-Gaussianity}",
    eprint = "1905.05697",
    archivePrefix = "arXiv",
    primaryClass = "astro-ph.CO",
    doi = "10.1051/0004-6361/201935891",
    journal = "Astron. Astrophys.",
    volume = "641",
    pages = "A9",
    year = "2020"
}

@article{Lyth:1999ty,
    author = "Lyth, David H.",
    title = "{Constraints on TeV scale hybrid inflation and comments on nonhybrid alternatives}",
    eprint = "hep-ph/9908219",
    archivePrefix = "arXiv",
    reportNumber = "LANCS-TH-9915",
    doi = "10.1016/S0370-2693(99)01089-8",
    journal = "Phys. Lett. B",
    volume = "466",
    pages = "85--94",
    year = "1999"
}

@article{Garcia-Bellido:1999xos,
    author = "Garcia-Bellido, Juan and Grigoriev, Dmitri Yu. and Kusenko, Alexander and Shaposhnikov, Mikhail E.",
    title = "{Nonequilibrium electroweak baryogenesis from preheating after inflation}",
    eprint = "hep-ph/9902449",
    archivePrefix = "arXiv",
    reportNumber = "IMPERIAL-TP-98-99-39, UCLA-99-TEP-7, UNIL-IPT-99-1",
    doi = "10.1103/PhysRevD.60.123504",
    journal = "Phys. Rev. D",
    volume = "60",
    pages = "123504",
    year = "1999"
}

@article{Krauss:1999ng,
    author = "Krauss, Lawrence M. and Trodden, Mark",
    title = "{Baryogenesis below the electroweak scale}",
    eprint = "hep-ph/9902420",
    archivePrefix = "arXiv",
    reportNumber = "CWRU-P11-99",
    doi = "10.1103/PhysRevLett.83.1502",
    journal = "Phys. Rev. Lett.",
    volume = "83",
    pages = "1502--1505",
    year = "1999"
}

@article{Freese:1990rb,
    author = "Freese, Katherine and Frieman, Joshua A. and Olinto, Angela V.",
    title = "{Natural Inflation with Pseudo - Nambu-Goldstone Bosons}",
    reportNumber = "FERMILAB-PUB-90-177-A",
    doi = "10.1103/PhysRevLett.65.3233",
    journal = "Phys. Rev. Lett.",
    volume = "65",
    pages = "3233--3236",
    year = "1990"
}

@article{Djouadi:2005gi,
    author = "Djouadi, Abdelhak",
    title = "{The Anatomy of electro-weak symmetry breaking. I: The Higgs boson in the standard model}",
    eprint = "hep-ph/0503172",
    archivePrefix = "arXiv",
    reportNumber = "LPT-ORSAY-05-17",
    doi = "10.1016/j.physrep.2007.10.004",
    journal = "Phys. Rept.",
    volume = "457",
    pages = "1--216",
    year = "2008"
}

@article{Degrassi:2012ry,
    author = "Degrassi, Giuseppe and Di Vita, Stefano and Elias-Miro, Joan and Espinosa, Jose R. and Giudice, Gian F. and Isidori, Gino and Strumia, Alessandro",
    title = "{Higgs mass and vacuum stability in the Standard Model at NNLO}",
    eprint = "1205.6497",
    archivePrefix = "arXiv",
    primaryClass = "hep-ph",
    reportNumber = "CERN-PH-TH-2012-134, RM3-TH-12-9",
    doi = "10.1007/JHEP08(2012)098",
    journal = "JHEP",
    volume = "08",
    pages = "098",
    year = "2012"
}

@article{Buttazzo:2013uya,
    author = "Buttazzo, Dario and Degrassi, Giuseppe and Giardino, Pier Paolo and Giudice, Gian F. and Sala, Filippo and Salvio, Alberto and Strumia, Alessandro",
    title = "{Investigating the near-criticality of the Higgs boson}",
    eprint = "1307.3536",
    archivePrefix = "arXiv",
    primaryClass = "hep-ph",
    reportNumber = "CERN-PH-TH-2013-166, FTUAM-13-20, IFT-UAM-CSIC-13-081, IFUP-TH",
    doi = "10.1007/JHEP12(2013)089",
    journal = "JHEP",
    volume = "12",
    pages = "089",
    year = "2013"
}

@article{ATLAS:2025eii,
    author = "Aad, Georges and others",
    collaboration = "ATLAS, CMS",
    title = "{Highlights of the HL-LHC physics projections by ATLAS and CMS}",
    eprint = "2504.00672",
    archivePrefix = "arXiv",
    primaryClass = "hep-ex",
    reportNumber = "ATL-PHYS-PUB-2025-018, CMS-HIG-25-002",
    month = "4",
    year = "2025"
}

@article{Ai:2025cpj,
    author = "Ai, Xiaocong and others",
    title = "{New physics search at the CEPC: a general perspective}",
    eprint = "2505.24810",
    archivePrefix = "arXiv",
    primaryClass = "hep-ex",
    doi = "10.1088/1674-1137/ae1194",
    journal = "Chin. Phys. C",
    volume = "49",
    pages = "123108",
    year = "2025"
}

@article{FCC:2025lpp,
    author = "Benedikt, M. and others",
    collaboration = "FCC",
    title = "{Future Circular Collider Feasibility Study Report: Volume 1, Physics, Experiments, Detectors}",
    eprint = "2505.00272",
    archivePrefix = "arXiv",
    primaryClass = "hep-ex",
    reportNumber = "CERN-FCC-PHYS-2025-0002",
    doi = "10.1140/epjc/s10052-025-15077-x",
    journal = "Eur. Phys. J. C",
    volume = "85",
    number = "12",
    pages = "1468",
    year = "2025"
}

@article{Arnowitt:1962hi,
    author = "Arnowitt, Richard L. and Deser, Stanley and Misner, Charles W.",
    title = "{The Dynamics of general relativity}",
    eprint = "gr-qc/0405109",
    archivePrefix = "arXiv",
    doi = "10.1007/s10714-008-0661-1",
    journal = "Gen. Rel. Grav.",
    volume = "40",
    pages = "1997--2027",
    year = "2008"
}

@article{Bezrukov:2014ipa,
    author = "Bezrukov, Fedor and Rubio, Javier and Shaposhnikov, Mikhail",
    title = "{Living beyond the edge: Higgs inflation and vacuum metastability}",
    eprint = "1412.3811",
    archivePrefix = "arXiv",
    primaryClass = "hep-ph",
    doi = "10.1103/PhysRevD.92.083512",
    journal = "Phys. Rev. D",
    volume = "92",
    number = "8",
    pages = "083512",
    year = "2015"
}

@article{Bezrukov:2014bra,
    author = "Bezrukov, Fedor and Shaposhnikov, Mikhail",
    title = "{Higgs inflation at the critical point}",
    eprint = "1403.6078",
    archivePrefix = "arXiv",
    primaryClass = "hep-ph",
    reportNumber = "CERN-PH-TH-2014-082",
    doi = "10.1016/j.physletb.2014.05.074",
    journal = "Phys. Lett. B",
    volume = "734",
    pages = "249--254",
    year = "2014"
}

@article{He:2014ora,
    author = "He, Hong-Jian and Xianyu, Zhong-Zhi",
    title = "{Extending Higgs Inflation with TeV Scale New Physics}",
    eprint = "1405.7331",
    archivePrefix = "arXiv",
    primaryClass = "hep-ph",
    doi = "10.1088/1475-7516/2014/10/019",
    journal = "JCAP",
    volume = "10",
    pages = "019",
    year = "2014"
}

@article{Xianyu:2014eba,
    author = "Xianyu, Zhong-Zhi and He, Hong-Jian",
    title = "{Asymptotically Safe Higgs Inflation}",
    eprint = "1407.6993",
    archivePrefix = "arXiv",
    primaryClass = "astro-ph.CO",
    doi = "10.1088/1475-7516/2014/10/083",
    journal = "JCAP",
    volume = "10",
    pages = "083",
    year = "2014"
}

\end{document}